\documentclass[aps, prx, notitlepage, twocolumn, longbibliography, footinbib,superscriptaddress]{revtex4-1}

\usepackage{amssymb}
\usepackage{amsmath}
\usepackage[colorlinks=true]{hyperref}
\usepackage{graphicx}
\usepackage{listings}
\usepackage{tabularx}
\usepackage{cancel}
\usepackage{comment}
\usepackage{hypcap}
\usepackage{subfigure}
\usepackage{color}
\usepackage[centerlast]{caption}
\usepackage{empheq}

\hypersetup{colorlinks=true, citecolor=blue, urlcolor=blue, linkcolor=blue}
\newcommand{\dbar}{d\hspace*{-0.08em}\bar{}\hspace*{0.1em}}

\newcommand{\magn}[1]{\left|#1\right|}
\newcommand{\op}[1]{\mathcal{O}_{#1}}

\newcommand{\bra}[1]{\left\langle#1\right|}
\newcommand{\ket}[1]{\left|#1\right\rangle}

\newcommand{\mvec}[1]{\mathbf{#1}}

\newcommand{\tr}[1]{\mathrm{Tr}\left[ #1 \right]}

\newcommand{\id}{\mathbb{I}}

\newcommand{\re}[1]{\mathrm{Re}[#1]}

\def\bea{\begin{eqnarray}}
\def\eea{\end{eqnarray}}
\def\nn{\nonumber}
\def\ba{\begin{array}}
\def\ea{\end{array}}
\def\Tr{\text{Tr}}
\def\nn{\nonumber}
\def\sgn{\text{sgn}}

\def\B{\textcolor{blue}}

\usepackage[shortlabels]{enumitem}

\graphicspath{{Figure/}}

\begin{document}

\date{\today}
\title{Entanglement phases in large-$N$ hybrid Brownian circuits with long-range couplings}

\author{Subhayan Sahu}
\email{subhayan@terpmail.umd.edu}
\affiliation{Condensed Matter Theory Center and Joint Quantum Institute,
Department of Physics, University of Maryland, College Park, MD 20742, USA}
\author{Shao-Kai Jian}
\affiliation{Martin A. Fisher School of Physics, Brandeis University, Waltham MA, USA}
\affiliation{Condensed Matter Theory Center and Joint Quantum Institute,
Department of Physics, University of Maryland, College Park, MD 20742, USA}
\author{Gregory Bentsen}
\affiliation{Martin A. Fisher School of Physics, Brandeis University, Waltham MA, USA}
\author{Brian Swingle}
\affiliation{Martin A. Fisher School of Physics, Brandeis University, Waltham MA, USA}
\affiliation{Condensed Matter Theory Center and Joint Quantum Institute,
Department of Physics, University of Maryland, College Park, MD 20742, USA}

\begin{abstract}
We develop solvable models of large-$N$ hybrid quantum circuits on qubits and fermions with tunable long-range power-law interactions and continuous local monitoring. These models provide analytical access to the entanglement phase diagram and error-correcting properties of many-body entangled non-equilibrium states generated by such dynamics. In one dimension, the long-range couplings are irrelevant for $\alpha>3/2$, where $\alpha$ is the power-law exponent, and the models exhibit a conventional measurement-induced phase transition between volume- and area-law entangled phases. For $1/2<\alpha<3/2$ the long-range couplings become relevant, leading to a nontrivial dynamical exponent at the measurement-induced phase transition. 
More interestingly, for $\alpha<1$ the entanglement pattern receives a sub-volume correction for both area-law and volume-law phases, indicating that the phase realizes a quantum error correcting code whose code distance scales as $L^{2-2\alpha}$. While the entanglement phase diagram is the same for both the interacting qubit and fermionic hybrid Brownian circuits, we find that long-range free-fermionic circuits exhibit a distinct phase diagram with two different fractal entangled phases.
\end{abstract}

\maketitle








\section{Introduction}

Modern quantum technologies facilitate increasingly detailed access to quantum phases of matter with complex patterns of many-body entanglement \cite{altmanQSimulationReview,Georgescu_2014}. In particular, long-range interactions decaying with distance as $r^{-\alpha}$, available in state-of-the-art experiments featuring Rydberg atoms, trapped ions, and neutral atoms in optical cavities, are capable of dramatically altering the dynamics of quantum information \cite{Eldredge_2017,Matsuta_2016,Tran_2020,Zhou2020,Chen_2019,Kuwahara_2020,Tran_2021,tran2021liebrobinson} and rapidly generating complex many-body entanglement~\cite{aikawa2012,saffman2010,Britton_2012,Yan_2013,Yao_2012,Douglas2015,monroe2021}. 
Our understanding of entanglement dynamics in generic strongly-interacting many-body systems is still under rapid development. A prime example is the recent discovery of robust phases of many-body entanglement which survive for exponentially long times in hybrid quantum circuits consisting of  scrambling unitary evolution interspersed with repeated local measurements~\cite{aharonov2000quantum,skinner2019measurement,li2018quantum,chan2019unitary,li2019measurement,gullans2020dynamical,choi2020quantum,zabalo2020critical,jian2020measurement,bao2020theory,gullans2020scalable,li2020conformal,szyniszewski2019entanglement,fan2020self,tang2020measurement,goto2020measurement,zhang2020nonuniversal,buchhold2021effective,jian2021yang,bao2021symmetry,sang2020measurement,lavasani2021measurement,lavasani2021topological,ippoliti2021fractal,ippoliti2021entanglement,lu2021entanglement,vijay2020measurement, bentsen2021measurementinduced,nahum2020measurement}. At low measurement rates initially local information is dynamically encoded in many-body entangled states which are robust to subsequent measurements, resulting in a volume-law-entangled phase stabilized by a dynamically-generated quantum error-correcting code (QECC)~\cite{choi2020quantum,bao2020theory,fan2020self,li2020statistical,li2021entanglement}. At higher measurement rates, the many-body entanglement is destroyed, leading to an area-law-entangled phase.
\begin{figure}
    \includegraphics[width =  0.9\columnwidth]{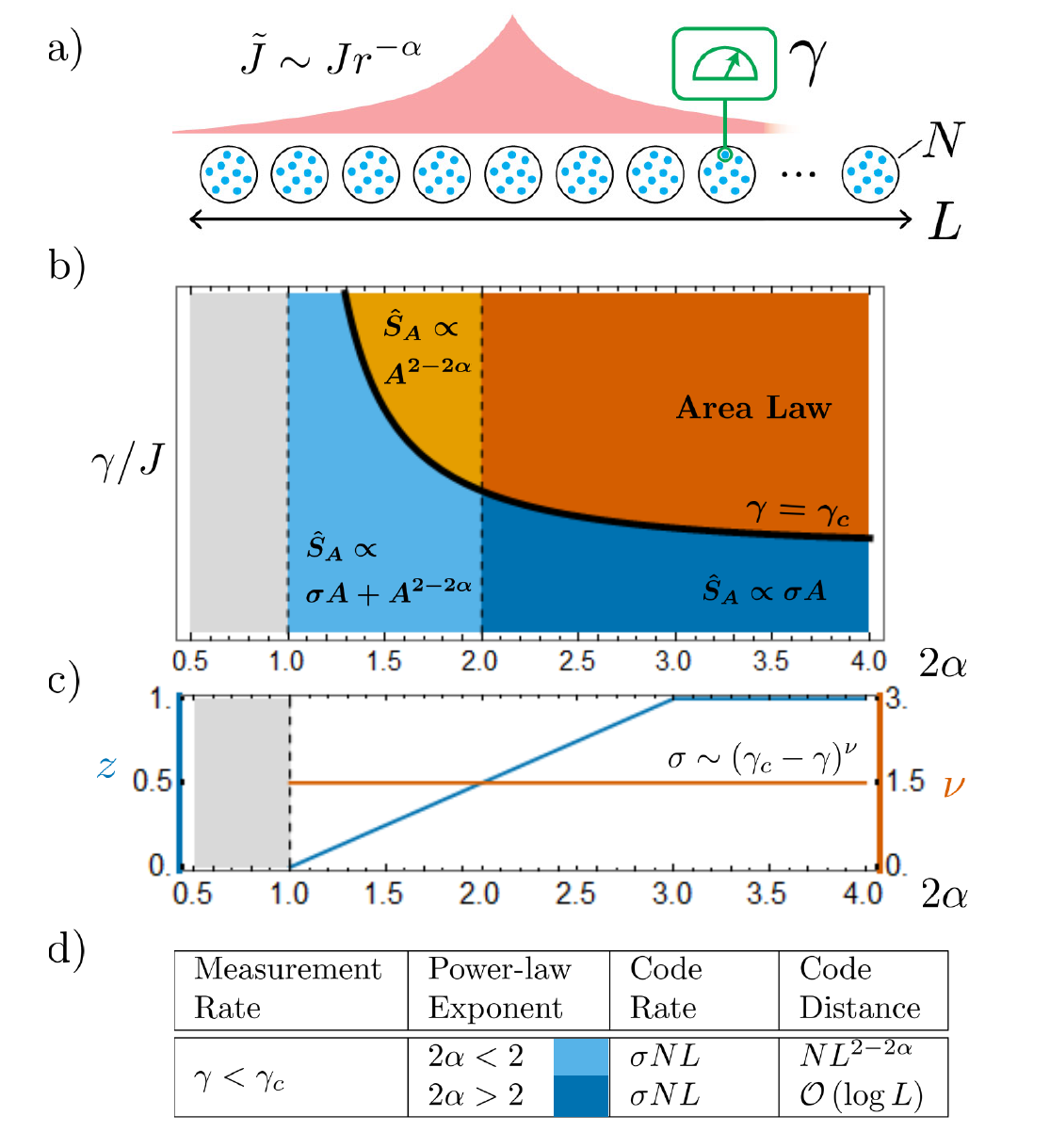} 
    \caption{(a) Monitored large-$N$ models with long-range interactions.
    (b) Entangled phases for Brownian spin and SYK$_4$ models 
    as a function of measurement rate $\gamma/J$ and long-range exponent $\alpha$. 
    $\hat{S}_A$ is the quasi-R\'enyi entropy of a contiguous subsystem of volume $A$. 
    (c) Dynamical critical exponent $z$ and domain wall tension critical exponent $\nu$ vs $\alpha$. 
    (d) Error-correcting properties of the measurement-induced phases at large $L$.}
    \label{fig:phases}
\end{figure}

It is highly desirable to develop theoretical tools to easily estimate entanglement properties of many-body states generated by strongly-interacting quantum dynamics. In this work we develop exactly solvable models composed of large-$N$ clusters of qubits or fermions in a 1D chain [Fig.~\ref{fig:phases}(a)], for which entanglement properties can be computed using path-integral techniques. 
In particular, these methods provide simple pen-and-paper calculations for the dynamics of many-body entanglement that can be immediately applied to problems of experimental interest.
We understand the entanglement phases in these models in terms of a replica-symmetry-breaking transition of a corresponding statistical mechanical system, and derive analytical expressions for entanglement entropies, code properties of the QECC phase, and critical properties of the phase transition. We summarize our primary findings in Figs.~\ref{fig:phases}(b-d), including a phase diagram, critical properties at the phase transition, and the error-correcting properties as a function of the measurement rate and the long-range exponent $\alpha$. 

Our analysis of these large-$N$ models leads to several new results and insights. 
First, our models go beyond the standard set of tools -- Clifford circuits, matrix product states, and exact diagonalization -- typically used to study entanglement dynamics, and are able to provide an analytical mean-field understanding of entanglement phases and measurement-induced phase transitions. Second, the analytical control afforded by our models allows us to derive new results, namely that long-range interactions generate novel sub-region entanglement structure, leading to a non-trivial QECC phase with a tunable sub-extensive code distance $L^{2-2\alpha}$ for $\alpha<1$, where $L$ is the number of clusters in the chain. The entanglement phase diagram we find is thus even richer than previously known~\cite{block2021measurementinduced,minato2021fate,muller2021measurementinduced}, and suggests a recipe for constructing QECCs with enhanced code distance. Third, the models we study are experimentally relevant as large-$N$ clusters naturally occur in cold atom experiments, including ensembles of atomic spins coupled uniformly to optical cavities  ~\cite{sorensen2002,baumann2010,leroux2010,bohnet2016} and ensembles of Rydberg atoms clustered within a single  blockade radius \cite{urban2009observation,gil2014spin}.
Fourth, our results highlight the crucial role played by interactions in stabilizing the volume-law phase. We demonstrate this point explicitly by studying a non-interacting circuit on fermions with long-range hopping and find two distinct fractal entangled phases~\cite{muller2021measurementinduced}, but no QECC phases. 
Lastly, although we focus here on the more easily computable R\'enyi entropies, we also expect similar calculations to allow for analytical calculations of von-Neumann entropies using an appropriate replica limit~\cite{jian2021phase}.

\section{Setup}\label{sec:setup}
\begin{figure}
    \includegraphics[width=0.9\columnwidth]{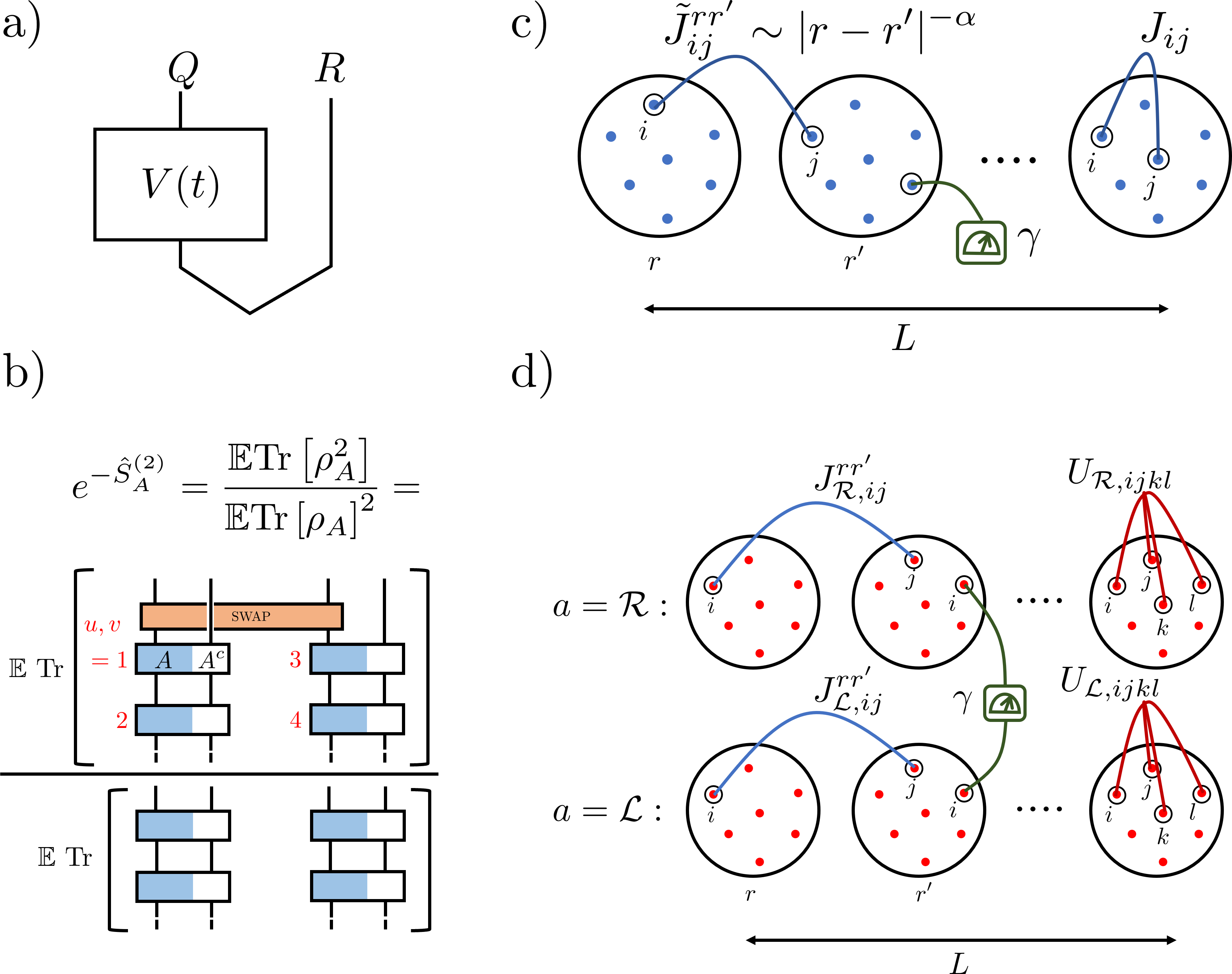}
    \caption{(a) System $Q$ is maximally entangled with reference $R$ and evolves under monitored dynamics $V(t)$. (b) Quasi R\'enyi-2 entropy represented as a quantum circuit. (c) Brownian qubit chain with $L$ clusters, each composed of $N$ qubits. (d) SYK$_4$ model with two independent chains of length $L$ coupled by measurement.}
    \label{fig:setup}
\end{figure}

Here we consider a system $Q$ of particles (qubits or fermions) arranged into a 1D chain of $L$ clusters, each containing a large number $N$ of particles as shown in Fig. \ref{fig:setup}. The particles are subjected to two competing dynamics: long-range Brownian unitary interactions $U(t)$ that rapidly generate entanglement leading to a volume-law phase, and continuous weak monitoring $M(t)$ that tends to destroy entanglement leading to an area-law phase. To probe the transition between these two phases, we maximally entangle the system $Q$ with a reference $S$ [Fig. \ref{fig:setup}(a)], and compute entanglement R\'enyi entropies $\hat{S}_A^{(n)}$ of system subregions $A \subset Q$. We shall show that these entropies can be readily computed using path-integral techniques, leading to straightforward pen-and-paper estimates for salient physical quantities including code properties and critical exponents.


For an analytical treatment, we focus on 
the quasi R\'enyi entropy~\cite{napp2020efficient} of the reduced density matrix $\rho_{A}$ of a contiguous subsystem $A$ of $Q$, 
$\hat{S}^{(n)}_{A} = -\log \frac{\mathbb{E}\tr{ \rho_{A}^{n}(t)}}{\mathbb{E}\tr{ \rho_{A}(t)}^{n}}$,
where $\mathbb E$ is a disorder average over measurement outcomes and circuit realizations ~\footnote{The relation between R\'enyi entropy, quasi R\'enyi entropy and the von Neumann entropy is explained in Appendix~\ref{appsec:relation_entropies}.}. The denominator is necessary for normalizing the state generated by the non-unitary evolution, and the ratio of averages is considered for ease of calculation over the averaged R\'enyi entropy which involves averaging the ratio of two multi-replica quantities.

Even though the quasi-entropy is distinct from the trajectory averaged R\'enyi entropies, one can show that trajectory averaged von-Neumann entropy is the $n\to 1$ limit of the quasi-R\'enyi entropy, $\hat{S}^{(n\to 1)}_{A} = \mathbb{E}{[S_{A}]}$ (for a proof, see Appendix~\ref{appsec:relation_entropies}). In this work, we focus on the computation of quasi-R\'enyi-2 entropy, $\hat{S}^{(2)}_{A}$.

$\hat{S}^{(2)}_{A}$ involves tracing over two copies of density matrix $\tr{\rho^{2}(t)}$ with $\rho(t) = V(t)\rho V^{\dagger}(t)$, which
we can interpret in terms of time evolution $\mathbf{V}(t) = V^{(1)}\otimes V^{(2)*}\otimes V^{(3)}\otimes V^{(4)*}$ on four replicas $r = 1,2,3,4$ \cite{bentsen2021measurementinduced,jian2021syk}. Here, $1,2$ ($3,4$) denote the first (second) R\'enyi replica, and $1,3$ ($2,4$) denote forward (backward) time-evolution. In the replicated Hilbert space, the quasi R\`enyi entropy can be expressed as a ratio of propagators, $\exp\left(-\hat{S}^{(2)}_{A}\right) = \frac{\langle\bra{\mathcal{S}_{A}}\mathbb{E} \mathbf{V}(t)\ket{\mathcal{I}}\rangle}{\langle\bra{\mathcal{I}}\mathbb{E} \mathbf{V}(t)\ket{\mathcal{I}}\rangle}$,
for appropriately defined initial and final states $|\mathcal{I}\rangle \rangle$, $|\mathcal{S}_{A}\rangle\rangle$ in the replicated Hilbert space, as demonstrated in Appendix~\ref{appsec:setup}. 
The numerator 
has twisted boundary conditions $|\mathcal{S}_{A}\rangle\rangle$ which swap replicas $1,3$ at the final time $t = T$ [Fig.~\ref{fig:setup}(b)], reflecting the SWAP test in the trace $\tr{\rho_A^2}$ \cite{Hastings_2010}.

We will construct analytically-tractable models for qubits and fermions for which the propagator $\mathbb{E}\mathbf{V}(t)$ can be expressed as a large-$N$ path integral with classical action $N I$ that is amenable to saddle point analysis. 
Consequently, the quasi R\'enyi entropy is simply proportional to the difference between the large-$N$ saddle-point actions with and without the SWAP boundary condition ~\cite{bentsen2021measurementinduced,jian2021syk,zhang2021universal}, $\hat{S}^{(2)}_{A} \propto N\left(I_{\text{SWAP}}-I\right).$

\section{Hybrid Brownian circuit on qubits}
We first consider a system of qubits $\mathcal{S}_{r,i,\alpha}$ residing in clusters $r$ and labeled by an intra-cluster index $i$ and spin component $\alpha=x,y,z$. 
During each time step $\Delta t$, the qubits evolve under a two-body scrambling unitary matrix $U(t) = \exp{[-i H(t) \Delta t / 2]}$ for half the duration, and under one-body weak measurement for the rest of the time. 
The unitary is generated by 
\bea
    \label{eq:brham}
    &H(t) = \sum_{\substack{r,i<j,\alpha\beta}} J_{ij\alpha \beta}(t) \mathcal{S}_{ri\alpha} \mathcal{S}_{rj\beta} \nn \\
    &+ \sum_{\substack{r'\neq r,ij\alpha\beta}} \tilde{J}^{rr^{\prime}}_{ij\alpha \beta}(t) \mathcal{S}_{ri\alpha} \mathcal{S}_{r^{\prime}j\beta},
\eea
with intra-cluster coupling $J_{ijuv}$, and inter-cluster coupling $\tilde{J}^{rr^{\prime}}_{ij\alpha \beta}$ between spins at site $r$ with spins at $r^{\prime}$ [Fig.~\ref{fig:setup}(c)]. 
A random 1-body operator at a given time, $\op{}(t) = \sum_{i,\alpha} n_i^{\alpha}(t) \mathcal{S}_{i\alpha}$ is coupled to an auxiliary qubit via the interaction,
\begin{equation}
    \label{eq:weakmeas}
    \exp{\left[- i \frac{\Delta t}{2} \op{}(t) \sigma^x_{\mathrm{aux}} \right] } \ket{\Psi}\ket{0}_{\mathrm{aux}},
\end{equation}
with $\ket{\Psi}$ being the state of the system. After this coupling, the auxiliary qubit is measured in the $\sigma^y_{\mathrm{aux}}$ basis, and only $+1$ results are post-selected. Under this dynamics, the system deterministically evolves with a non-unitary evolution operator $M(t)$, 
\begin{align}
    \label{eq:nonunitarym}
    \ket{\Psi} \to M(t) \ket{\Psi} = \left( 1 - \frac{1}{2} \op{} \Delta t - \frac{1}{8} \op{}^2 \Delta t^2 + \cdots \right) \ket{\Psi}.
\end{align}

The system evolves under a circuit constructed by stacking alternating layers of $U(t)$ and $M(t)$ gates, $V(t) \equiv \prod_{t = 0}^T M(t) U(t)$.

We consider Brownian Gaussian couplings and measurements with zero mean and variance, $\mathbb{E}\left[n(t) n(t')\right]\sim \delta_{tt^{\prime}}/\Delta t$, $\mathbb{E}\left[J(t) J(t')\right]\sim \delta_{tt^{\prime}}/(N\Delta t)$  and 
$\mathbb{E}\left[\tilde{J}(t) \tilde{J}(t')\right] \sim gJ|r_{1}-r_{2}|^{-2\alpha}\delta_{tt^{\prime}}/(N\Delta t)$~\cite{smsecII}. Here we have suppressed the indices of $n,J, \tilde{J}$ which should be considered to be independent and random, and the factor of $N$ is introduced for a meaningful large-$N$ limit.
We also take the continuum time limit $\Delta t \rightarrow 0$. This model allows us to make analytical progress in accessing the quasi-R\'enyi entropies.

\subsection{Path Integral Representation}

We now introduce multiple replicas $u,v\in\{1,2,3,4\}$ of the system, and use SWAP tests between the replicas to measure R\'enyi entropies \cite{daley2012measuring,islam2015measuring}. Here we want to represent the replicated circuit $\mathbf{V}(t) = V^{(1)}\otimes V^{(2)*}\otimes V^{(3)}\otimes V^{(4)*}$ as a path integral over collective replica fields.

Averaging over the random couplings $J,\tilde{J},n$ introduces inter-replica interaction, which in the large-N limit are mean-field couplings $G^{uv}_{r}\sim 1/N\sum_{i}\mathcal{S}^{u}_{ri}\cdot \mathcal{S}^{v}_{ri}$ between replicas $u,v$. We decouple the $G$ fields by introducing Hubbard-Stratonovich replica-fields
$iF_{r}^{uv}$. The averaged circuit $\mathbb{E}\mathbf{V}(t)$ can now be expressed as a path integral in these fields, $\mathbb{E}\mathbf{V}(t) = \int \mathcal{D}\left[iF_{r}^{uv}\right]\exp \left(- I\left[iF_{r}^{uv}\right]\right)$ with the action,
\bea \label{eq:actionreplicafield}
      & \frac{I\left[iF_{r}^{uv}\right]}N  =\sum_{r}\Big[\sum_{u<v}\int_{t}\big((-1)^{u+v+1}\sum_{r^{\prime}}\mathcal{J}_{rr^{\prime}}\left(iF_{r}^{uv}\right)\left(iF_{r^{\prime}}^{uv}\right)\nonumber
     \\&-\Gamma\left(iF_{r}^{uv} \right)\big)-\log \mathcal{K}_{r}\Big]. 
\eea
$\mathcal{K}_{r} \!=\! \tr{\exp\left(\int_{t} \sum_{u<v}\frac{-iF_{r}^{uv}}{(\mathcal{S}+1)^{2}}(\mathcal{S}^{u}_{r}\cdot \mathcal{S}^{v}_{r})\right)}$ is the spin propagator which determines the bulk theory. The derivation of this path integral representation directly follows~\cite{bentsen2021measurementinduced}, and is detailed in the Appendix~\ref{appsec:spinmodelpathintegral}.

Since the interaction between the replica fields in different sites is derived from the power-law interaction among the physical qubits in the circuit, the renormalized interaction strength is also long-range, $\mathcal{J}_{rr^{\prime}}\sim |r-r^{\prime}|^{-2\alpha}$ (see detailed derivation in Appendix~\ref{appsec:renormalized_locality}). 

For specific matrix elements like $\langle\bra{\mathcal{S}_{A}}\mathbb{E}\mathbf{V}(t)\ket{\mathcal{I}}\rangle$, the spin propagator $\mathcal{K}_{r}$ has to be evaluated with fixed boundary condition instead of the trace.

The action (\ref{eq:actionreplicafield}) describes the dynamics of $4L$ spins $\mvec{S}_r^u$ interacting via Heisenberg couplings $\mvec{S}_r^u \cdot \mvec{S}_r^v$. Because these coupling terms are manifestly symmetric under global $\mathrm{SU}(2)$ rotations, the action $I = I[iF_r^{uv}]$ is also $\mathrm{SU}(2)$ invariant. In fact, because the interaction terms are separable $\sum_r \ln \mathcal{K}_r$ in the space coordinate $r$, the action $I$ is invariant under all \emph{local} $\mathrm{SU}(2)$ rotations generated by the total spin operators $\mvec{S}_r^{\mathrm{Tot}} = \sum_u \mvec{S}_r^u$ within each cluster $r$. At the same time, the boundary conditions generated by the EPR pairs and SWAP operator force the system to form $\mathrm{SU}(2)$ spin singlets at times $t = 0,T$. Together, these facts constrain the dynamics within each cluster $r$ to live entirely in the spin-singlet subspace $\mvec{S}_r^{\mathrm{Tot}} = 0$ for each $r = 1,\ldots,L$. Thus, each cluster $r$ supports a single replica qubit or r-bit spanned by the states $\ket{\uparrow}_r,\ket{\downarrow}_r$ \cite{bentsen2021measurementinduced}.

The underlying $\mathrm{SU}(2)$ symmetry significantly simplifies the problem. For $n = 2$ the propagator $\mathcal{K}_r$ in each two-dimensional r-bit subspace simplifies to
\begin{align}
    \label{eq:rbitpropagator}
    \mathcal{K}_{r} = \bra{\psi_T} \exp{\left[ \frac{1}{2} \int_0^T dt \ \big( \phi_r(t) \sigma^x_r + \Theta_r(t) \sigma^z_r \big) \right]} \ket{\psi_0} e^{B T / 2}
\end{align}
where $\sigma^{x,z}_r$ are the $2 \times 2$ Pauli matrices acting on the r-bit $\ket{\uparrow}_r,\ket{\downarrow}_r$ subspace. The fields $\phi_r(t),\Theta_r(t)$ are linear combinations of the mean fields
\begin{align}\label{eq:2field_def}
    \phi_r &= \frac{2}{3\sqrt{3}} \left(i F^{12}_r + i F^{34}_r - i F^{14}_r - i F^{23}_r \right) \nonumber \\
    \Theta_r &= \frac{2}{9} \sum_{u<v} i F^{uv}_r - \frac{2}{3} \left( i F^{13}_r + i F^{24}_r \right)
\end{align}
and where terms proportional to the identity within the r-bit subspace have been collected into the term
\begin{equation}
    B = \frac{2}{9L} \sum_{r, u < v} i F^{uv}_r.
\end{equation}
The time- and space-dependent bulk fields $\phi_r(t),\Theta_r(t)$ encode the relevant mean-field dynamics of the r-bits in each cluster $r$. In general these fields must execute nontrivial motions in the bulk in order to satisfy the non-equal boundary conditions $\ket{\psi_0},\ket{\psi_T}$. By contrast, the remaining fields in the action appear simply as quadratic Gaussian fields and may therefore be trivially integrated out of the path integral, leading to the effective action
\bea \label{eq:2fieldPathIntegralmain}
& \mathbb{E}\tr{ \rho_{A}^{2}(t)} \ \mathrm{or} \ \mathbb{E}\tr{ \rho_{A}(t)}^{2} = \int \mathcal{D}\phi\mathcal{D}\Theta \exp{\left[-N I[\phi,\Theta]\right]},\nonumber\\
& I[\phi,\Theta] = \sum_r \int_{0}^{T} dt \bigg[\sum_{r'}\frac{27\mathcal{J}_{rr'}}{16}\left(\phi_{r}\phi_{r'} - 3\Theta_{r}\Theta_{r'} \right) \nonumber\\ &+ \Theta_r \left(1+\frac{9\gamma}{\hat{J}_0}\right)\bigg]- \sum_r \ln \mathcal{K}_{r}.
\eea
with the propagator $\mathcal{K}_{r}$ given in (\eqref{eq:rbitpropagator}), and where we have dropped additive constant terms. 

\subsection{Replica permutation symmetry breaking}

The action \eqref{eq:actionreplicafield} is invariant under the replica symmetry group $ G = (S_2\times S_2)\rtimes \mathbb{Z}_{2}$,
where the two inner $S_2 \cong \mathbb{Z}_2$ denote permutations of the forward and backward replicas $ 1 \leftrightarrow 3$ and $2 \leftrightarrow 4$ \cite{bentsen2021measurementinduced,nahum2020measurement}.
The outer $\mathbb{Z}_2$ in the semidirect product is generated by time-reversal $\mathcal{T}$ on four replicas followed by exchange of even and odd replicas $1 \leftrightarrow 2$, $3 \leftrightarrow 4$. 
The boundary states for the entropy $\hat{S}_{A}^{(2)}$ explicitly break $ 1 \leftrightarrow 3$ (or equivalently $2 \leftrightarrow 4$) symmetry. 


Saddle point analysis of the simplified bulk action (\ref{eq:2fieldPathIntegralmain}) in the mean field limit reveals a phase transition at $\gamma_{c} = \frac{J}{9}\left(1+2g\zeta(2\alpha)\right)$ with the Riemann Zeta function $\zeta(\alpha) \equiv \sum_{r=1}^\infty \frac1{r^\alpha}$. The order parameter field  is $\phi$, as defined in (\ref{eq:2field_def}) as $\sim \frac{2}{3\sqrt{3}}\left(iF_{12}+iF_{34}-iF_{14}-iF_{23}\right)$. For $\gamma > \gamma_c$ the saddle point is $\phi=0$, while for $\gamma < \gamma_c$, $\phi$ is non-zero and comes in a pair, $\phi \propto \pm \sqrt{\gamma_{c}-\gamma}$. 
The replica permutation $1\leftrightarrow 3$ is equivalent to $\phi\leftrightarrow-\phi$ symmetry, which is spontaneously broken for $\gamma < \gamma_c$. The details of the saddle point computation are presented in Appendix~\ref{appsec:spinmodelsaddle}.

Near the critical point the dynamics is governed entirely by the small fluctuations of the symmetry-breaking field $\phi_r(t)$. In the following we expand the action $I[\phi,\Theta]$ in fluctuations of this small parameter to obtain an effective long-range $\phi^4$ field theory for the system (see Appendix~\ref{appsec:spinmodelsaddle}) 
\bea \label{eq:phiaction}
&    \frac{I_{\text{eff}}}{N}=\int_{t,r} [ -  \phi \partial_{t}^{2}\phi -b\int^{\prime}_{s} \frac{\phi_{r}\phi_{s}}{|r-s|^{2\alpha}}- \phi \partial_{r}^{2}\phi -\frac{\delta}{2}\phi^{2}+\frac{\  \phi^{4}}{4}].
\eea
For $2\alpha>1$ this has a phase transition effected by the mass term $\delta \propto \gamma_{c}-\gamma$, with $\delta >(<) 0$ being the symmetry-broken (symmetric) phase, and the $\mathbb{Z}_{2}$ phase transition occurring at $\delta \to 0$. 
The long-range term appears with a regulated integral: $\int^{\prime}_{s}$ to be read as $\int_{\mathbb{R}\backslash(r-\varepsilon,r+\varepsilon)}ds$, with an ultraviolet cut-off $\varepsilon$. 
For $b\neq 0$, this is the long wavelength theory of an anisotropic long range Ising model \cite{Fisher1972LongRange,Sak1973,Paulos_2016} in 2D, where the interaction is long-range along space and short-range along time. 
For $2\alpha<1$ the power law contribution diverges, so $J$ must be scaled with $L$ to take the thermodynamic limit: the system behaves as a single all-to-all cluster with $NL$ qubits without any volume to area-law transition. 
There is however a phase transition in the purification times for the system and its parts~\cite{bentsen2021measurementinduced}. 

For $2\alpha\!>\!3$ the long-range term is irrelevant and the transition is governed by the short-ranged fixed point. For $2\alpha \!<\!3$, the underlying Ising model is anisotropic with a non-trivial dynamical critical exponent,
\bea \label{eq:dynamical}
 &   z = \begin{cases}
        1 & 2\alpha > 3 \\
        \frac{2\alpha - 1}{2} &  1 < 2\alpha < 3.
    \end{cases}
\eea
In this regime we have two distinct correlation lengths $\xi_{t}\sim \delta^{-\frac{1}{2}}$ and $\xi_{r}\sim \delta^{-\frac{1}{2z}}$ corresponding to the time-like and space-like directions, respectively. 

\begin{figure}
    \includegraphics[width=\columnwidth]{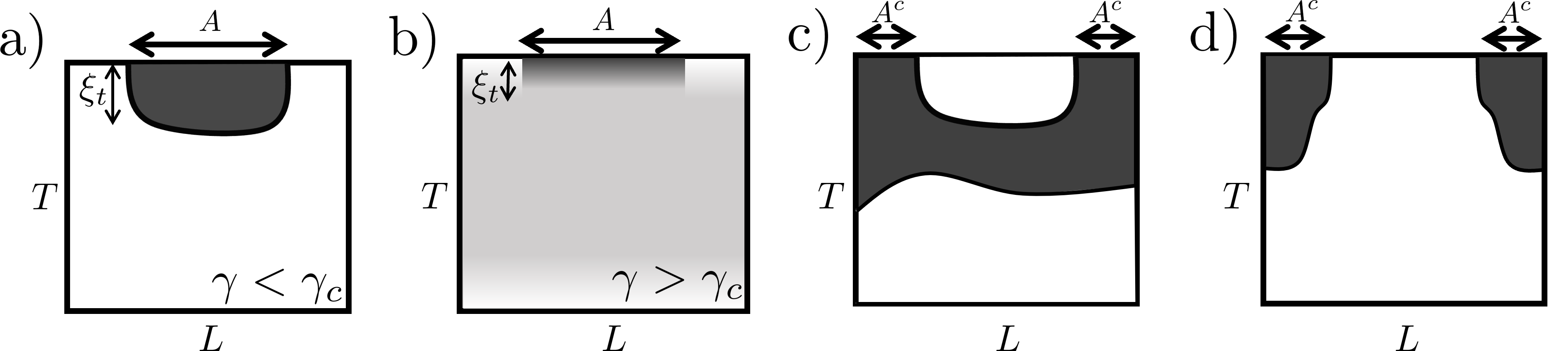}
    \caption{Domains and domain walls in the anisotropic Ising model corresponding to the quasi-entropy $\hat{S}_A^{(2)}$ of a small subregion $A$ in the ferromagnetic phase (a, black $\phi > 0$ and white $\phi < 0$ are symmetry-broken domains separated by a domain wall), and the paramagnetic phase $\phi = 0$ (b, light gray). The entropy of the complementary subregion $A^c$ corresponds to one of two possible competing domain-wall configurations (c,d).} 
    \label{fig:domainwall}
\end{figure}

\subsection{Entanglement phases}

Using the effective action \eqref{eq:phiaction} we now calculate the quasi entropy of $\rho_A(t)$ for a contiguous region $A \subset Q$. 
The twisted boundary condition in $I_{\text{SWAP}}$ corresponds to pinning $\phi > 0$ within the subregion $A$ at the future time boundary $t = T$, and pinning $\phi < 0$ on all other boundaries~[Fig.~\ref{fig:domainwall}(a,b)]. 
In the symmetry-broken phase $\delta > 0$ the bulk organizes into domains separated by a domain wall [Fig.~\ref{fig:domainwall}(a)] such that the bulk field is positive $\phi > 0$ (black) within a temporal correlation length $\xi_{t}$ of the final time-boundary of the subregion $A$ and negative $\phi < 0$ (white) throughout the remainder of the bulk.

In the symmetric phase $\delta<0$ only the trivial bulk saddle point $\phi = 0$ [Fig.~\ref{fig:domainwall}(b), light gray] contributes, and therefore this pinning effect is only relevant very close to the final time boundary .
For $ A\ll L$ the excess energy $I_{\text{SWAP}} - I$ of this configuration compared to the configuration without the twisted boundary condition is simply given by the energy cost of the power-law interaction acting along a spatial slice of height $\xi_{t}$:
\bea
\label{eq:symmentropy}
    & J \xi_{t} \int^{\prime}_{r\in A, s\in A^{c}} dr ds \frac{1}{|r-s|^{2\alpha}}  
    \sim J\xi_{t} A^{2-2\alpha}  +\text{const.}
\eea

In the symmetry broken phase $\delta > 0$, the twisted boundary condition at $t = T$ supports a bulk domain wall with spatial extent $A$ and time-like height $\xi_{t}$.
The space-like part of the domain wall with spatial extent $A$ has a domain wall tension
$\sigma_{r} \sim \xi_{t}\left(\phi/\xi_{t}\right)^{2}\propto \delta^{\frac{3}{2}}$. 
The time-like part of the domain wall also contributes an energetic term (sub-extensive in $A$) arising from the power-law just like~(\ref{eq:symmentropy}). Combining these results together we have,
\bea \label{eq:entropyInteracting}
    \frac{\hat{S}^{(2)}_{A}}{N} \sim \begin{cases}
        cJ \xi_{t}A^{2-2\alpha} &\text{ for }\gamma > \gamma_c
        \\
        J\left(\sigma_{r}A + c\xi_{t}A^{2-2\alpha}\right) &\text{ for }\gamma <  \gamma_c.
    \end{cases}
\eea

In the absence of power-law terms, we obtain $1/N$ logarithmic corrections to the entropy in the symmetry-broken phase for the local model, which arise from the fluctuations of the domain wall, similar to the capital wave theory predictios~\cite{li2020statistical}, which is discussed in Appendix~\ref{appsec:symmbroken}.

We emphasize that the power-law correction to the entropy in the volume-law (symmetry-broken) phase is a novel result, which leads to enhanced error-correcting properties for long-range hybrid circuits. 

  
\subsection{Error correcting properties} 

$\hat{S}^{(2)}_R$ in the volume-law phase $\gamma < \gamma_c$ can be understood to be the `rate' of the QECC, which refers to the amount of logical information of $R$ that is encoded in $Q$ and protected from `errors' due to measurements with a Code Rate $\sim \hat{S}^{(2)}_{R}\sim \sigma N L$.
The mutual information $\hat{I}^{(2)}(A:R) = \hat{S}^{(2)}_{A}+\hat{S}^{(2)}_{R}-\hat{S}^{(2)}_{A^{c}}
$ between a subregion A and the reference R is related to the contiguous `code distance', which refers to the size of the largest contiguous subsystem of $Q$ whose deletion would not spoil the encoded information of $R$ \cite{li2020statistical}\footnote{This statement comes with two important caveats. First, the theorem proved in \cite{li2020statistical} rigorously applies only to 
Clifford circuits (stabilizer states), whereas here our Brownian circuit elements clearly take the quantum state outside the Clifford group. Second, to make meaningful information-theoretic arguments one must ultimately work with the disorder-averaged von Neumann entropy $-\mathbb{E}[\tr{\rho_A \log \rho_A}]$}. From \eqref{eq:entropyInteracting} we have $\hat{S}^{(2)}_{A} \sim N\left(\sigma A + c\xi_{t}A^{\upsilon}\right) +\mathcal{O}\left(1\right) $ and $\hat{S}^{(2)}_{R} \sim N\sigma L +\mathcal{O}\left(1\right)$, where $\upsilon \equiv 2-2\alpha$. The quasi entropy $\hat{S}^{(2)}(A^{c})$ is the minimum of two configurations in Fig.~\ref{fig:domainwall}(c,d)~\cite{li2020statistical,li2021entanglement},
\bea
 &   \hat{S}^{(2)}_{A^{c}} \sim  \min \left\{\hat{S}^{(2)}_{A}+\hat{S}^{(2)}_{R},N\left(\sigma (L-A) + c\xi_{t}(L-A)^{\upsilon}\right)\right\}. \nn
\eea
The cross-over between the two occurs for a critical subregion size $A^{*}\sim  \frac{1}{2\sigma}L^{\upsilon}+\mathcal{O}(N^{-1})$. Thus, for $A<A^{*}$ we have $\hat{I}^{(2)}(A:R)\approx 0$, and $N A^{*}$ can be identified as a power-law code distance,
\bea 
\text{`Code Distance'} \sim N L^{\upsilon} \text{ for }\gamma<\gamma_c.
\eea
The distance can be tuned with the long-range exponent $\alpha$, and is sub-linear but scales favorably with $L$ for $2\alpha < 2$. For $2\alpha > 2$, the code distance is $1/N$ suppressed  and scales as $\log L$ in our model (see Appendix~\ref{appsec:symmbroken}). 

\section{Monitored SYK chain}
We now turn to the study of the effects of long-range couplings on the fermionic monitored Brownian Sachdev-Ye-Kitaev (SYK) chain circuit introduced in~\cite{jian2021syk}. This allows us to separately consider the effects of the long-range coupling and the interactions, which highlight the role of interactions on the entanglement properties of the states generated by hybrid circuits. 

\subsection{Model}
The setup  contains a left ($\mathcal{L}$) and a right ($\mathcal{R}$) chain with $L$ clusters of $N$ Majorana  fermions each~\cite{chen2017competition,song2017strongly,jian2021syk} [Fig.~\ref{fig:setup}(d)], with chain undergoing intermittent unitary evolution and monitoring. The unitary evolution is generated by inter-cluster long-range two-fermion  of strength $J_{ij}^{rr^{\prime}}\sim |r-r^{\prime}|^{-\alpha}$ and on-site four-fermion interaction of strength $U_{ijkl}^{r}$, which are both independent Brownian variables for each chain. 

The Brownian Gaussian random couplings in the unitary part is defined by the parameters,
\bea\label{eq:random}
& \mathbb{E} \left[J_{a,ij}^{r_{1}r_{2}}(t_1) J_{a',ij}^{r^{\prime}_{1}r^{\prime}_{2}}(t_2)\right] = \frac{J_{rr^{\prime}}}{2N} \frac{\delta_{t_{1}t_{2}}}{\Delta t}  \delta_{aa'} \delta^{r_{1},r_{1}'} \delta^{r_{2},r_{2}'}, \\
& \mathbb{E} \left[U_{a,j_1...j_q}^{r}(t_1) U_{a',j_1...j_q}^{r'}(t_2)\right] = \frac{2^{q-2}(q-1)! U}{N^{q-1}} \frac{\delta_{t_{1}t_{2}}}{\Delta t}  \delta_{aa'} \delta^{r,r'}. \nn
\eea.

We focus on the study of the free-fermion limit $U \to 0$ which allows us to consider the effects of long-range hopping and an on-site interaction separately. The $\mathcal{L}$ and $\mathcal{R}$ chains are coupled by a inter-chain parity measurement for each site, described by Kraus operators, $\{ M_1^{r,i}, M^{r,i}_2\} \!=\! \left\{   \pi^-_{r,i} + \sqrt{1-s^2} \pi^+_{r,i}  ,s \pi^+_{r,i} \right\}$,     
where $\pi^\pm_{x,i}=\frac{1}{2} (1 \mp i 2 \psi_{r,L,i} \psi_{r,R,i})$. $s$ denotes the measurement strength.


The measurement part for the flavor $i$ at the site $x$ can be cast into (where $u$ refers to the replica index),
\bea
 &&\sum_{\nu=1}^2 M_\nu^{x,i} \otimes M_\nu^{x,i\dag} \otimes M_\nu^{x,i} \otimes M_\nu^{x,i\dag} \nn\\
 &=& \left(1 - \frac{s^2}2 \sum_{u=1}^4 \pi^{+,u}_{x,i} + s^4 \otimes_{u=1}^4\pi^{+,u}_{x,i} \right) \nn \\
    &\approx& \exp   \left( - \frac{s^2}2 \sum_{u} \pi^{+,u}_{x,i}\right) \nn\\
    & = & \exp  \frac{\delta t \gamma}{2}  \sum_u i \psi_{x,L,i}^u \psi_{x,R,i}^u,
\eea
where we have used the relation $\pi^+_{x,a,j} + \pi^-_{x,a,j} = 1$ and also introduced $u = 1,...,4$ to denote the four copies of the tensor product. To derive the above equation, we assume $ s \ll 1$ and keep orders up to $O(s^2)$. $s$ is chosen as~\cite{wiseman1996quantum} $s = \sqrt{\gamma \Delta t}$, with $\Delta t \rightarrow 1$ keeping $\gamma$ fixed. In the last line we introduce $ \gamma = s^2/\delta t $, and when the continuum limit is taken, $\gamma$ is kept fixed. All the constants are neglected because they will not affect the dynamics. The effect of monitoring every Majorana species $i$ at every site $x$ is described by
\bea \label{eq:measurement}
 \exp  \left(\frac\gamma2 \int dt  \sum_{x,u,i} i \psi_{x,L,i}^u \psi_{x,R,i}^u\right),
\eea 
where we implicitly sum over all infinitesimal time steps to arrive at the time integral for a time evolution.



\subsection{Path integral and saddle point computation}

Due to the large-$N$ structure in both the unitary and the monitoring part~(\ref{eq:measurement}), we can introduce the bilocal field,
\bea
    G_{ab,r}^{uv}(t,t') = \frac1N \sum_j \psi_{r,a,j}^u(t) \psi_{r,b,j}^v(t'),
\eea
to rewrite the Majorana field with the help of the following identity
\bea
    1 &=& \int D\Sigma \exp \int dt_1 dt_2 \Big[ - \frac{N}2 \Sigma^{uv}_{ab,r}(t_1,t_2) \Big( G_{ab,r}^{uv}(t_1,t_2) \nn\\&&- \frac1N  \sum_i \psi_{r,a,i}^u(t_1) \psi_{r,b,i}^v(t_2)\Big) \Big],
\eea
where $\Sigma_{ab,r}^{uv}(t,t')$ is the self-energy. It is a standard approach for the SYK model, which is then generalized to the four contours with monitoring part~\cite{jian2021syk}. With a slight modification that replace the nearest-neighbor hopping to a power-law hopping, the large-$N$ action in the replica space reads
\bea \label{eq:large-N}
- \frac{I}N &=&  \sum_{r} \Big[ \frac12 \Tr \log \left( (-1)^{u+1}  \partial_t - \Sigma_{r} \right)- \frac12 \int_{t,t'}  \Sigma_{ab,r}^{uv} G_{ab,r}^{uv}  \nn \\
 	 && + \int_{t,t'} \delta(t-t') \Big[ \frac{(-1)^{u+v+1}}{4} \delta_{ab} \Big( \sum_{r^{\prime}} J^{rr'} G_{ab,r}^{uv} G_{ab,r'}^{uv} \nn \\&& + \frac{U}{2q} (2G_{ab,r}^{uv})^{q} \Big) + \frac{i \gamma}2 G_{LR,r}^{uu} \Big] \Big],
\eea
where $u,v = 1,...,4$ denote the four contours, and $\int_{t,t'} \equiv \int dt dt'$. The summations over $a,b$ and $u,v$ are implicit. Note that the model can be generalized to arbitrary graph with a modification on the hopping term in the second line of~(\ref{eq:large-N}).

Saddle-point analysis can be straightforwardly applied to the large-$N$ action. The Schwinger-Dyson equation resulted from~(\ref{eq:large-N}) reads
\bea
    [G_r^{-1}]^{uv}_{ab} &=& (-1)^{u+1} \delta^{uv} \delta_{ab}\partial_t - \Sigma_{ab,r}^{uv}, \nn \\
    \Sigma_{ab,r}^{uv} &=& \delta(t-t') \Big[ \frac{ (-1)^{u+v+1} \delta_{ab}}{2}  \Big( \sum_{r'} J_{rr'} 2G_{ab,r'}^{uv} \nn \\   + &&U(2 G_{ab,r}^{uv})^{q-1} \Big) + i \gamma \delta^{uv} \frac{\delta_{aL}\delta_{bR} -\delta_{aR}\delta_{bL}}{2}\Big]. \nn
\eea
For a homogeneous solution in real space, i.e., $G_{ab,r}^{uv} = \bar G_{ab}^{uv}$ and $\Sigma_{ab,r}^{uv} = \bar \Sigma_{ab}^{uv}$, the Schwinger-Dyson equation is simplified to be
\bea
    [\bar G^{-1}]^{uv}_{ab} &=& (-1)^{\alpha+1} \delta^{uv} \delta_{ab}\partial_t - \bar \Sigma_{ab}^{uv}, \nn \\
    \bar \Sigma_{ab}^{uv} &=& \delta(t-t') \Big[ \frac{ (-1)^{u+v+1} \delta_{ab}}{2}  \Big(  \hat J 2\bar G_{ab}^{uv}  \nn \\&&+ U(2 \bar G_{ab}^{uv})^{q-1} \Big) + i \gamma \delta^{uv} \frac{\delta_{aL}\delta_{bR} -\delta_{aR}\delta_{bL}}{2}\Big], \nn
\eea
where it is convenient to define $\hat J = J \zeta(2\alpha) $ and $\tilde{\gamma} = \gamma/\hat{J}$ and $\tilde{U} = U/\hat{J}$.

To get the solution, we focus on two contours, $u, v = 1,2$, because the boundary condition in $\Tr(\rho)^2$ is to connect 1 to 2 and connect 3 to 4 separately. 
According to Ref.~\cite{jian2021syk}, the saddle point solution can be obtained by replacing $J$ to $\hat J $,
\bea \label{eq:SYKq_saddle}
 \bar G (t_1,t_2) = \begin{cases} \frac{e^{- \frac{\hat J + U \lambda^{q-2}}2 |t_{12}|}}{2} \Big[\sgn(t_{12}) \sigma^z - \lambda i \sigma^y + \frac{\tilde \gamma \tau^y}{1+\tilde U \lambda^{q-2}} \Big] \\ \qquad \qquad \qquad \qquad \qquad \qquad \qquad \qquad \tilde \gamma < 1, \\
\\
    \frac{e^{- \frac{\gamma |t_{12}|}2 }}{2} \left( \sgn(t_{12})\sigma^z +  \tau^y  \right) \\ \qquad \qquad \qquad \qquad \qquad \qquad\qquad \qquad \tilde \gamma \ge 1, 
	\end{cases} \nn 
\eea
where $t_{12} \equiv t_1 -t_2$ is the time difference, $\tilde \gamma \equiv \gamma/ \hat J$, $\tilde U \equiv U/\hat J$ and Pauli matrix $\sigma $ ($\tau$) acts on 1 and 2 contours ($L$ and $R$ chains). The parameter $\lambda$ is given by 
\bea \label{eq:lambda}
    (1-\lambda^2)(1+ \tilde U \lambda^{q-2})^2 = \tilde \gamma^2.
\eea
The solution on $3,4$ contours is the same, consistent with the boundary condition without twist operators. We will discuss the saddle-point solution in more details in the next section.

\subsection{The non-interacting case}

We consider the free fermion limit, $U=0$. For Brownian randomness, it is legitimate to assume the Green functions are strictly local $G_{ab,r}^{uv}(t, t) \equiv G_{ab,r}^{uv}(t) $ and antisymmetric $G^{uv}_{ab,r}(t) = -G^{vu}_{ba,r}(t)$~\cite{saad2019semiclassical} so for $U=0$ the action can be written as
\bea \label{eq:trace}
&& - \frac{I}N  = \frac12 \Tr \log \left( S \partial_t + \Sigma_{r} \right) + \int  \frac12 \Tr\Big[ \Sigma_{ab,x} G_{ba,x} + \nn\\&&\sum_{r'}\frac{J_{rr'}}{4}  G_{ab,r} S G_{ba,r'} S + i \frac{\gamma}2 G_{LR,r} \Big], 
\eea
where $ S^{uv} = (-1)^{u} \delta^{uv}$ and the trace in the second line is over the contour indices. The theory is invariant under $O(2) \times O(2)$ transformation, i.e., 
\bea
    G_{ab,x} \rightarrow O^{-1} G_{ab,x} O, \quad O^T O = 1, \quad O^T S O = S, 
\eea
where $O$ acts identically on the left and right chains. (Without the coupling between the left and the right chains, $\mu=0$ , the action is invariant under two $O(2) \times O(2)$ for the left and the right chains, respectively, $G_{ab,x} \rightarrow O^{-1}_a G_{ab,x} O_b$, where $O^T_a O_a = 1, O^T_a S O_a = 1$.) The rotational symmetry is generated by $\gamma_{(13)}$ and $\gamma_{(24)}$, $\gamma_{(ij)}^{uv} =  \delta^{iu} \delta^{jv} -\delta^{ju} \delta^{iv}$, i.e.,
\bea
    O = e^{\theta_{13} \gamma_{(13)} + \theta_{24} \gamma_{(24)}},
\eea
where $\theta_{13}$ denotes the rotation angle between the contour $1$ and $3$, and $\theta_{24}$ denotes the rotation angle between the contour $2$ and $4$.

The saddle point solution~(\ref{eq:SYKq_saddle}) for $U=0$ and $\tilde \gamma < 1$ spontaneously breaks the {\it relative} rotational symmetry. For the noninteracting case, $U = 0$, $ \lambda = \sqrt{1 -\tilde \gamma^2} $. There is one Goldstone mode generated by applying the broken-symmetry generator $\gamma_- \equiv \gamma_{(13)} - \gamma_{(24)} $, i.e.,  
\bea \label{eq:Goldstone}
\delta G_{aa,r}(t) &=& e^{-\theta_r(t) \gamma_-} \bar G_{aa}(t,t) e^{\theta_r(t) \gamma_-} - \bar G_{aa}(t,t) \nn \\&&\approx \sqrt{1- \tilde \gamma^2} \theta_r(t) (\gamma_{(14)} + \gamma_{(23)}),
\eea
where $\theta_r(t)$ denotes the Goldstone mode. In contrast, when $\tilde \gamma > 1$, this $O(2)$ symmetry is unbroken and the replicated theory is in the gapped phase. 

\subsection{Landau-Ginzburg theory}

We thus found that the theory for quasi-R\'enyi entropy for the free-fermionic case ($U = 0$) is invariant under the replica symmetry group $(O(2)\times O(2))\rtimes \mathbb{Z}_{2}$, where the two $O(2)$ transformations rotate the $1,3$ and $2,4$ contours, respectively. This continuous symmetry (in contrast to the discrete symmetry for the qubits) is spontaneously broken for $\gamma \!<\! \hat{J}$, resulting in a Goldstone mode $\theta_r$ ~\cite{goldstone} corresponding to the relative $O(2)$ rotation angle at site $r$. The effective theory of the Goldstone mode can be derived~\cite{jian2021syk},
\bea
    \frac{I_{\text{eff}}}{N} = \frac{\rho}2 \sum_k \int_\Omega  \left( \frac{\Omega^2}{\gamma^2}  + (1-\epsilon_k) \right) |\theta_k(\Omega)|^2,
\eea
where $\theta_k = \frac1{\sqrt{L}}\sum_r \theta_r e^{-i k r} $ is the Fourier transform of the lattice site, and 
\bea
    \epsilon_k \equiv \frac1{\zeta(2\alpha)} \sum_{r=1}^\infty \frac{\cos kr}{r^{2\alpha}},
\eea
resulted from the power-law hopping.
Similar results have been obtained in Ref.~\cite{zhang2021universal}. 
Notice that $\epsilon_{k=0} = 1$ consistent with $\theta_k(\Omega)$ being a Goldstone mode.
The stiffness $\rho = \hat J ( 1- \tilde \gamma^2) $ vanishes at $\tilde \gamma  = 1$, indicating that the transition occurs at $\gamma = \hat J$. 

The long-range couplings lead to a nontrivial dynamical exponent as before~(\ref{eq:dynamical}). In the long wavelength limit, the kinetic term of Goldstone mode can be approximated by
\bea \label{eq:kinetic}
    1- \epsilon_k \approx 
\begin{cases} \frac{\zeta(2(\alpha-1))}{\zeta(2\alpha)}\frac{k^2}2 & \alpha> \frac32 \\
    - \frac{\Gamma(1-2\alpha) \sin \pi \alpha}{\zeta(2\alpha)} k^{2\alpha - 1} &  \frac12 < \alpha < \frac32
\end{cases}.
\eea
When $\alpha> 3/2$, the kinetic term is the same as a local hopping amplitude~\cite{jian2021syk}, which leads to $z=1$.
On the other hand, when $1 < \alpha < 3/2$, the leading term in kinetic term $k^{2\alpha-1}$ has a power less then two, leading to the same nontrivial dynamical exponent as in the Brownian spin model ~(\ref{eq:dynamical}).

\subsection{Entanglement Phases}

\begin{figure}
    \includegraphics[width = 0.9 \columnwidth]{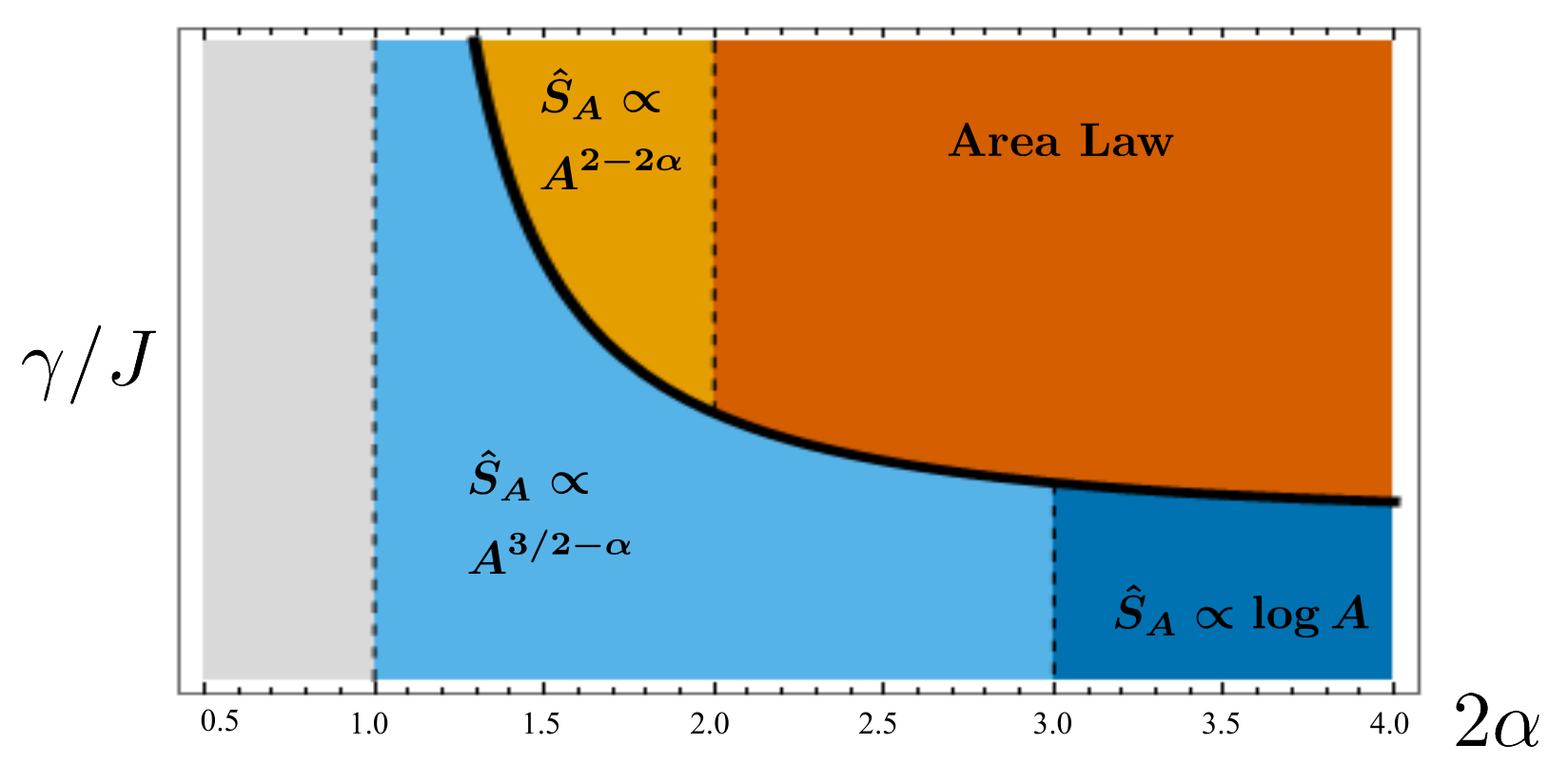}
    \caption{Phases of the long-range monitored SYK$_2$ free-fermion model. For $\alpha > 3/2$ the long-range couplings are irrelevant and the model has the usual area-law phase (dark orange) above the transition (black line) and a logarithmic entanglement phase driven by vortex formation (dark blue) below the transition. For $1/2 < \alpha < 3/2$ the symmetry-broken phase (light blue) has fractal entanglement, with scaling intermediate between area-law and volume-law. Additionally, for $\alpha < 1$ the normal phase also gives way to fractal entanglement (light orange), but with a different scaling than the symmetry-broken phase.}
    \label{fig:phases_nonint}
\end{figure}

In the symmetry-broken phase, the boundary condition pins the angle $\theta=\pi/2$ in subsystem $A$, and $\theta = 0$ in $A^c$~\cite{zhang2021universal}. This is equivalent to creating half-vortices at the left and right boundaries of $A$, such that the quasi entropy can be mapped to the correlation function of a vortex creation operator $\varphi$. The vortex creation operator is the dual to the $O(2)$ field via, $\partial_t \theta \sim \partial_x \varphi$~\cite{zhang2021universal, fradkin2013field}. Then the quasi entropy scaling can be obtained from the scaling dimension of the vortex creation operator. Since the scaling dimension of $\theta_x(t)$ is $[\theta] = \frac{1-z}2$, the scaling dimension of $\varphi_x(t)$ is $[\varphi] = [\theta] + z - 1 = \frac{z-1}2$. The subsystem quasi entropy in the symmetry-broken phase has the scaling form
\bea
    \frac{\hat{S}^{(2)}_A}{N} \propto \begin{cases} \log A & \alpha > 3/2 \\
    A^{3/2-\alpha} & 1/2 < \alpha < 3/2
    \end{cases}.
\eea
For $\alpha>3/2$ we have the well-known logarithmic free energy for vortices when the long-range coupling is irrelevant. The free energy of vortices is proportional to the stiffness and its critical exponent is simply $\nu = 1$ for all $\alpha > 1/2$.

When we turn on interactions $U>0$, the replica symmetry is reduced to a discrete group $(C_{4} \!\times\! C_{4})\!\rtimes\! \mathbb{Z}_{2}$. There is consequently a $Z_4$ symmetry-breaking transition which reproduces the same entanglement and error-correcting phase diagram as the spin model as shown in  Fig.~\ref{fig:phases}(b). The interacting case is detailed in Appendix~\ref{appsec:syk}.

\section{Discussion}
In this paper we introduced and studied analytically-solvable models at large-$N$ which exhibit entanglement phase transitions in the presence of measurements and tunable long-range couplings. We considered power-law couplings in large-$N$ Brownian circuits in continuous time for both spins and Majorana fermions and for both interacting and non-interacting models. We discovered that the long-range couplings non-perturbatively modify the entanglement structure of the usual measurement-induced phase diagram with volume and area law phases. In the interacting models, we found that the quantum states generated by these circuit possess good quantum code properties, namely a finite code rate and a power-law code distance. In the non-interacting models (but still with power-law hoppings), we identified multiple fractally-entangled phases. We demonstrated the generic nature of these properties by showing the same behavior for both spin and fermionic models.

The phase diagram in Fig.~\ref{fig:phases} can be readily generalized to Brownian chains in higher dimensions, and demonstrates an entanglement transition for all $2\alpha>d$, with non-trivial dynamical exponent $z = (2\alpha-d)/2$ for $d<2\alpha<d+2$. The subextensive correction arising from~(\ref{eq:symmentropy}) is $A^{2d-2\alpha}$. Interestingly, in the long-range interacting models with $2\alpha<2$, we found a non-perturbative code distance that is a tunable sub-extensive power of system size. A fixed power-law scaling has also been numerically observed for local hybrid Clifford circuits \cite{li2020statistical}, which has been attributed to the quenched disorder in the models \cite{li2021entanglement}. It will be interesting to investigate the effect of disorder on the power-law code distance in the long-range case. 


Since all the phases and transitions discussed here were obtained in the large-$N$ limit, it interesting to ask whether perturbative corrections in $1/N$ match the well-known anomalous dimensions for the 2D Ising model or for the percolation transition, or if they constitute distinct universality classes. 
It will also be interesting for future studies to study the R\'enyi entropy for general $n$ and the entanglement entropy at $n \rightarrow 1$.


\section{Acknowledgement}

SS is supported in part by AFOSR under Award FA9550-17-1- 0180. SKJ and BGS are supported by the Simons Foundation via the It From Qubit Collaboration. GB is supported by the DOE GeoFlow program (DE-SC0019380). BGS is supported in part by the AFOSR under grant number FA9550-19-1-0360. SS and SKJ have contributed equally to the paper.


%

\onecolumngrid
\begin{appendix}

\section{Setup of the R\'enyi entropy calculation}\label{appsec:setup}
In the main part of the paper we focus on a 1-dimensional lattice of size L with periodic boundary conditions, however the results can be readily extended to higher dimensions. For the spin model, each spin operator is labeled as $\mathcal{S}_{ri\alpha}$, with $i$ and $\alpha$ denoting the intra-cluster label and spin direction respectively, while $r$ refers to the position label of the cluster in the chain, and a similar labeling applies to the fermions. 

The dofs in Q (vectors in the Hilbert space $\mathcal{H}$) undergo a non-unitary random evolution $V(t)$. The randomness comes from the Brownian nature of the unitary evolution that we consider in our setup. Under this evolution, a density matrix (when viewed as vector in the doubled Hilbert space $\ket{\rho}\rangle$) evolves as, $\ket{\rho}\rangle \to V(t)\otimes V^{*}(t)\ket{\rho}\rangle$, which generates an unnormalized state because of the non-unitary evolution. We want to study the entanglement properties of the subsystem $A$ for this state, a particular diagnostic of which is the R\'enyi-n entropy of the normalized reduced density matrix $\rho_{A}$,

\begin{align}
    S^{(n)}_{A} = \frac{1}{1-n}\mathbb{E}\log \frac{\tr{ \rho_{A}^{n}(t)}}{\tr{ \rho_{A}(t)}^{n}},
\end{align}
where $\mathbb{E}$ refers to the averaging over the realizations of the random circuit $V(t)$.

Estimating the actual R\'enyi-2 entropy averaged over the randomness requires taking a non-trivial replica limit since this is an average of a ratio of two multi-replica quantities. Instead, we will calculate the quasi R\'enyi-2 entropy \cite{napp2020efficient}, which is the ratio of the averages, 
\begin{align}
    \hat{S}^{(n)}_{A} = \frac{1}{1-n}\log \frac{\mathbb{E}\tr{ \rho_{A}^{n}(t)}}{\mathbb{E}\tr{ \rho_{A}(t)}^{n}}
\end{align}
This quantity, while distinct from the disorder averaged R\'enyi-n entropy, is easier to calculate. In this work we focus on estimating $\hat{S}^{(2)}_{A}$, for which it is convenient to consider the dynamics in 4 replicas of the Hilbert space $\mathcal{H}^{(4)}$. We use a $1,2,3,4$ notation: $1,2$ ($3,4$) denote the first (second) replica, and $1,3$ ($2,4$) denote the forward (backward) evolution. In $\mathcal{H}^{(4)}$ the time evolution is given by $\mathbf{V}(t) = V^{(1)}(t)\otimes V^{(2)*}(t)\otimes V^{(3)}(t)\otimes V^{(4)*}(t)$, and the normalization factor is given by $\mathcal{N}(\rho) = \sqrt{\langle\bra{\mathcal{I}}\mathbf{V}(t)\ket{\rho}\rangle\ket{\rho}\rangle}$. To define $\ket{\mathcal{I}}\rangle \in \mathcal{H}^{(4)}$, we first define two normalized states,
\begin{align}
    \ket{+}\rangle &\sim \sum_{ab = 0,1}\ket{aabb}\rangle\\
    \ket{-}\rangle &\sim \sum_{ab = 0,1}\ket{abba}\rangle,
\end{align}
which leads to the definition, $\ket{\mathcal{I}}\rangle = \bigotimes_{i,r}\ket{\mathcal{+}}\rangle$. This definition ensures that the normalization factor is the usual norm of a state. In our setup, the system dofs in $Q$ are initially maximally entangled with $NL$ qubits in a reference $R$. The corresponding initial state for Q in $\mathcal{H}^{(4)}$ is  $\ket{\rho}\rangle\otimes\ket{\rho}\rangle = \ket{\mathcal{I}}\rangle$. 

We further define a SWAP state in $\mathcal{H}^{(4)}$ as,
\begin{align}
    \ket{\mathcal{S}_{A}}\rangle = \bigotimes_{i,r\in A}\ket{-}\rangle \bigotimes_{i,r\in A^{c}=Q-A}\ket{+}\rangle,
\end{align}
and the quasi R\'enyi-2 entropy is given by,
\begin{align}
    \exp\left(-\hat{S}^{(2)}_{A}\right) = \frac{\langle\bra{\mathcal{S}_{A}}\mathbb{E} \mathbf{V}(t)\ket{\mathcal{I}}\rangle}{\langle\bra{\mathcal{I}}\mathbb{E} \mathbf{V}(t)\ket{\mathcal{I}}\rangle}.
\end{align}

The quasi-entropy can be simulated in a quntum experiment with extra classical post-processing and no extra quantum resources over estimating the usual R\'enyi entropy~\cite{bentsen2021measurementinduced}.

\subsection{Relation between the quasi R\'enyi, R\'enyi, and von Neumann entropies}\label{appsec:relation_entropies}

One can show that the R\'enyi entropy and the quasi R\'enyi entropy can be treated as a part of the same family of generalized entropic quantities. Consider the generalized function,
\bea
\chi^{(nm)}_{A} = \frac{1}{m(1-n)}\log\frac{\mathbb{E}\left(\tr{\rho_{A}^{n}}\right)^{m}}{\mathbb{E}\left(\tr{\rho_{A}}\right)^{nm}}
\eea
One can show,
\begin{align}
    \chi^{(nm)}_{A} \big|_{m\to 0} &= S^{(n)}_{A} \\
    \chi^{(nm)}_{A} \big|_{m\to 1} &= \hat{S}^{(n)}_{A}. 
\end{align}

Intriguingly, one can also extract the von Neumann entropy directly from the quasi R\'enyi entropy. First let us define the normalized density matrix, $\tilde{\rho}_{A} = \rho_{A}/\tr{\rho_{A}}$. The probability of each trajectory is given by $\tr{\rho_{A}}$. The trajectory averaged von-Neumann entropy is given by,
\bea
\mathbb{E}[{S_{A}}] = -\mathbb{E}\big[\tr{\rho_{A}}\tr{\tilde{\rho}_{A}\log{\tilde{\rho}_{A}}}\big].
\eea
We find that the trajectory averaged von-Neumann entropy is the $n\to 1$ limit of the quasi-R\'enyi entropy,
\bea
\hat{S}^{(n\to 1)}_{A} = \mathbb{E}{[S_{A}]}.
\eea
Let us sketch this proof. 
\begin{align*}
    \mathbb{E}{[S_{A}]} &= -\mathbb{E}\big[ \tr{\rho_{A}\log{\rho_{A}}}-\tr{\rho_{A}}\log{\tr{\rho_{A}}}\big]\\
    & = -\partial_{n}\mathbb{E}\big[ \tr{\rho_{A}^{n}}-\left(\tr{\rho_{A}}\right)^{n} \big]\bigg|_{n\to 1}\\
    &= -\partial_{n}\left[\frac{\mathbb{E}\tr{\rho^{n}}}{\mathbb{E}\left(\tr{\rho_{A}}\right)^{n}}\right]\bigg|_{n\to 1} \text{ (using $\mathbb{E}\tr{\rho_{A}}= 1$)}\\
    &= \frac{1}{1-n}\log\frac{\mathbb{E}\tr{\rho^{n}}}{\mathbb{E}\left(\tr{\rho_{A}}\right)^{n}}\bigg|_{n\to 1} \\
    &= \hat{S}^{(n\to 1)}_{A}.
\end{align*}
Hence we find that although the quasi R\'enyi and the averaged R\'enyi entropies are different replica limits of a generalized entropy function, a particular replica limit of the quasi R\'enyi entropy can access the physical von Neumann entropy of the state generated by the hybrid circuit.

\section{Path integral representation of the replicated dynamics}\label{appsec:spinmodelpathintegral}

In this section of the Appendix we show how to derive the spin path integral for the replicated dynamics and derive Eqs.~\ref{eq:actionreplicafield},5 and 6 in the main text.


\subsection{Integrating out disorder and spin path integral}

The Brownian disorders are explicitly given by,
\bea
    \label{eq:brownianvar}
    &\mathbb{E}\left[n_i^{\alpha}(t) n_{j}^{\beta}(t')\right] = \frac{\gamma}{(\mathcal{S}+1)^2} \frac{\delta_{tt'}}{\Delta t/2} \delta_{i j} \delta^{\alpha \beta}, \nonumber\\
    &\mathbb{E}\left[J_{i j\alpha \beta}(t) J_{k l\mu \rho}(t')\right] = \frac{J}{N (\mathcal{S}+1)^{4}} \frac{\delta_{tt'}}{\Delta t/2} \delta_{i k} \delta_{j l} \delta_{\alpha \mu} \delta_{\beta \rho}, \\
     &\mathbb{E}\left[ \tilde{J}^{r_{1}r_{2}}_{ij\alpha \beta}(t) \tilde{J}^{r^{\prime}_{1}r^{\prime}_{2}}_{k l\mu \rho}(t') \right] = \frac{g J_{r_{1}r_{2}}}{N (\mathcal{S}+1)^{4}} \frac{\delta_{tt'}}{\Delta t/2}\delta_{r_{1}r_{1}^{\prime}}\delta_{r_{2}r_{2}^{\prime}} \delta_{ik}\delta_{jl} \delta_{\alpha \mu} \delta_{\beta \rho}. \nn
\eea

We want to write down a path integral expression for the averaged circuit evolution in $n$ replicated copies of the physical system, which includes $2n$ copies of the circuit (counting the time-reversed copies). This quantity, \begin{align}
\mathbf{V}(t) = \mathbb{E}\left[ V^{(1)}(t) \otimes V^{(2)*}(t)\otimes \ldots V^{(2n-1)}(t)\otimes V^{(2n)*}(t)\right],
\end{align}

can give us access to the $n$-th R\'enyi entropies, and in particular, the $n = 2$ result corresponds to the R\'enyi-2 entropy that we consider in this paper. Let $u,v$ denote the replica index. The time evolution operator in any given copy is generated by an interaction Hamiltonian and a non-unitary evolution generated by a weak measurement followed by post-selection, as described in the main text. Schematically this can be expressed as, 
\begin{align}\label{eq:evolutionschematic}
    V^{(u)}(\Delta t)\sim \exp\left[- \frac{i\Delta t}{2}\left(H^{(u)}(t)-i\mathcal{O}^{(u)}(t)\right)\right],
\end{align}
where the $H^{(u)}$ and $\mathcal{O}^{(u)}$ are the Hamiltonian and the measured operator (in replica $u$) respectively. These operators are taken to be Brownian in our model, which implies that the time evolution is uncorrelated in the time direction. However the randomness is same for each replica, which implies that in $\mathbf{V}$ different replicas become correlated when the disorder is integrated away, using the Gaussian nature of the Brownian variables defined in~Eq. 3 of the main text. One way to derive this is by expanding the evolution in each time step in the product in ~(\ref{eq:evolutionschematic}) to second order in $\Delta t$, and collecting terms like $\mathbb{E}\left[H^{(u)}H^{(v)}\right]$, $\mathbb{E}\left[H^{(u)}H^{(u)}\right]$ or $\mathbb{E}\left[\mathcal{O}^{(u)}\mathcal{O}^{(v)}\right]$ together. This procedure is discussed in detail in the Appendix B of \cite{bentsen2021measurementinduced}. Exponentiating back, one obtains that $\mathbf{V} (\Delta t)\sim \exp\left(-N I_{n}(t)\Delta t\right)$. The variances of the random terms have been scaled in a way that the exponential comes with a prefactor of $N$. Dropping some constant terms (which arise from the intra-replica terms) $I_{n}(t)$ is given by,

\begin{align}\label{eq:action_def}
    I_{n}(t)  &= \Bigg[\sum_{\substack{r\\u<v}}(-1)^{u+v}\frac{J}{4(\mathcal{S}+1)^{4}}\left(\frac{1}{N}\sum_{i}\mathcal{S}_{ri}^{u}\cdot \mathcal{S}_{ri}^{v}\right)^{2}\nonumber \\
    & +\sum_{\substack{r\neq r^{\prime}\\u<v}}(-1)^{u+v}\frac{g J_{rr^{\prime}}}{2(\mathcal{S}+1)^{4}}\left(\frac{1}{N}\sum_{i}\mathcal{S}_{ri}^{u}\cdot \mathcal{S}_{ri}^{v}\right) \left(\frac{1}{N}\sum_{i}\mathcal{S}_{r^{\prime}i}^{u}\cdot \mathcal{S}_{r^{\prime}i}^{v}\right) \nonumber\\
    &-\sum_{\substack{r\\u<v}}(-1)^{u+v}\frac{\gamma}{2(\mathcal{S}+1)^{2}}\left(\frac{1}{N}\sum_{i}\mathcal{S}_{ri}^{u}\cdot \mathcal{S}_{ri}^{v}\right)\Bigg].
\end{align}
In this expression the spin operator $\mathcal{S}$ refers to the spin operator at time $t$.

We can now stack  $\mathbf{V}_{u}(\Delta t)$ at different times by repeating sequence of the Brownian interactions and measurements, insert resolutions of the identity between each layer, and take the limit $\Delta t \rightarrow 0$ with $T$ fixed to express $\mathbb{V}(t)$ as a path integral over $2nN$ unit-norm $\mathrm{SO}(3)$ spins $\mvec{\mathcal{S}}_{ri}^{u}$, using spin coherent states as the basis. The completeness relation for the coherent states for a single spin is given by,
\begin{equation}
    \id = \int \frac{2\mathcal{S}+1}{4\pi}d\mvec{\Omega}_{i}\ket{\mvec{\Omega}_{i}}\bra{\mvec{\Omega}_{i}}.
\end{equation}
To turn the spins into coherent states, we use the upper symbols for single spin-$S$ Pauli operators \cite{Klauder1985},
\begin{align}
    \mathcal{S}_{i\alpha} &= \int \frac{2\mathcal{S}+1}{4\pi}d\mvec{\Omega}_{i} \ket{\mvec{\Omega}_{i}}\bra{\mvec{\Omega}_{i}}(\mathcal{S}+1) \Omega_{i\alpha}\\
    \left(\mathcal{S}_{i\alpha}\right)^{2} &= \int \frac{2\mathcal{S}+1}{4\pi}d\mvec{\Omega}_{i} \ket{\mvec{\Omega}_{i}}\bra{\mvec{\Omega}_{i}}\left[(\mathcal{S}+1)\left(\mathcal{S}+\frac{3}{2}\right)\left(\Omega_{i\alpha}\right)^{2}-\frac{\mathcal{S}+1}{2}\right],
\end{align}
and introduce a measure for the coherent spin states in the path integral,
\begin{equation}
    \mathcal{D}\Omega_{ri}^{u} = \prod_{t_{n}}\frac{2\mathcal{S}+1}{4\pi}d\mvec{\Omega}^{u}_{r,i,t_{n}}\langle\mvec{\Omega}^{u}_{r,i,t_{n+1}}|\mvec{\Omega}^{u}_{r,i,t_{n}}\rangle,
\end{equation}
with implicit time dependent terms.

In terms of the coherent states $\mathbb{V}(t)$ is given by a path integral,

\begin{align}
    &\mathbf{V}(t) = \mathbb{E}\left[ V \otimes V_{\mathcal{T}} \otimes \cdots \otimes V \otimes V_{\mathcal{T}}\right]  = e^{- N I[\mvec{\Omega}]} \nonumber \\
    I[\mvec{\Omega}] = \int_{0}^{T}dt &\Bigg[\frac{J}{4}\sum_{\substack{r\\u<v}}(-1)^{u+v}\left(\frac{1}{N}\sum_{i}\mvec{\Omega}_{ri}^{u}\cdot \mvec{\Omega}_{ri}^{v}\right)^{2}\ \nonumber\\
    & +\sum_{\substack{r\neq r^{\prime}\\u<v}}(-1)^{u+v}\frac{g J_{rr^{\prime}}}{2}\left(\frac{1}{N}\sum_{i}\mvec{\Omega}_{ri}^{u}\cdot \mvec{\Omega}_{ri}^{v}\right) \left(\frac{1}{N}\sum_{i}\mvec{\Omega}_{r^{\prime}i}^{u}\cdot \mvec{\Omega}_{r^{\prime}i}^{v}\right) \nonumber\\
    &-\sum_{\substack{r\\u<v}}(-1)^{u+v}\frac{\gamma}{2(S+1)^{2}}\left(\frac{1}{N}\sum_{i}\mvec{\Omega}_{ri}^{u}\cdot \mvec{\Omega}_{ri}^{v}\right)\Bigg]
\end{align}

To decouple the non-linear interactions in $\mvec{\Omega}_{ri}^{u}\cdot\mvec{\Omega}_{ri}^{v}$, we introduce Hubbard Stratonovich type fields which couple different replicas $F_{uv}^{(r)}$ and $G_{uv}^{(r)}$, satisfying the operator identity,
\begin{equation}
    1 = \left(\prod_{r,u<v}\int \mathcal{D}F^{uv}_{r} \mathcal{D}G^{uv}_{r}\right)\exp\left[iN\int_{0}^{T}dt F^{uv}_{r}\left(G^{uv}_{r}-\frac{1}{N}\sum_{i}\mvec{\Omega}_{ri}^{u}\cdot \mvec{\Omega}_{ri}^{v}\right)\right].
\end{equation}

The action can thus be re-written as,
\begin{align}\label{eq:mipt_loc_action}
    I[F,G,\mvec{\Omega}]&=  \int dt\Bigg[\sum_{\substack{r\\u<v}}\frac{(-1)^{u+v}J}{4}\left(G^{uv}_{r}\right)^{2}+\sum_{\substack{r\neq r^{\prime}\\u<v}}\frac{(-1)^{u+v}g J_{rr^{\prime}}}{2}\left(G^{uv}_{r}\right) \left(G^{uv}_{r^{\prime}}\right) -\sum_{\substack{r\\u<v}} \frac{(-1)^{u+v}\gamma}{2}\left(G^{uv}_{r}\right)\nonumber\\
    &-\sum_{\substack{r\\u<v}}\left(iF^{uv}_{r}G^{uv}_{r}\right)+\sum_{\substack{r\\u<v}}iF^{uv}_{r}\left(\frac{1}{N}\sum_{i}\mvec{\Omega}_{ri}^{u}\cdot\mvec{\Omega}^{v}_{ri}\right)\Bigg]
\end{align}

Note that~(\ref{eq:mipt_loc_action}) is applicable in any general dimensions, where the position label $r\in \mathbb{Z}^{d}$ can be identified as a $d$ dimensional vector, with $L$ being the linear size along each dimension. 

\subsection{Field theory with periodic boundary condition}

We can further simplify~(\ref{eq:mipt_loc_action}) by assuming periodic boundary condition and going to momentum space. We also assume that the interaction $J_{rr^{\prime}}$ is translationally invariant and even, i.e. $J_{rr^{\prime}}\sim J_{|r-r^{\prime}|}$.

We consider discrete d-dimensional cubic lattice in the space $\mathbb{Z}^{ d}$ with $L$ being the linear extent of the cube. In the limit $L\gg 1$, the momentum space domain is $k \in \{0,2\pi\}^{ d}$. We use the following schematic definitions,
\begin{align}
    &G_{k} = \sum_{r\in \mathbb{Z}^{ d}} e^{-ik\cdot r}G_{r} \nonumber \\
    &G_{r} = \frac{1}{(2\pi)^{d}}\int_{0}^{2\pi}\cdots \int_{0}^{2\pi}dk_{1}\cdots dk_{d} e^{ik\cdot r}G_{k} \equiv \int \dbar^{d} k e^{ik\cdot r} G_{k}\nonumber\\
    &\frac{1}{(2\pi)^{d}}\sum_{r\in \mathbb{Z}^{ d}} e^{ik\cdot r} = \delta (k) \ ,\ \int \dbar^{d} k e^{ikr} = \delta_{r,0}.
\end{align}
We also introduce the notation, 
\begin{align}
    \Omega_{r}^{uv} = \frac{1}{N}\sum_{i}\mvec{\Omega}_{ri}^{u}\cdot\mvec{\Omega}_{ri}^{v}.
\end{align}
In momentum space, we can rewrite Eq. \ref{eq:mipt_loc_action} can be rewritten as (dropping the constant terms),
\begin{align}\label{eq:mipt_loc_action_mom}
    I[F,G,\mvec{\Omega}]&=  \int dt \int \dbar^{d} k\sum_{\substack{u<v}}\Bigg[\frac{(-1)^{u+v}}{4}\left(\hat{J}_{k} G_{k}^{uv}G_{-k}^{uv}-4\pi\gamma G_{k}^{uv}\delta(k) \right)-iF_{k}^{uv}G_{-k}^{uv}+iF_{k}^{uv}\Omega^{uv}_{-k}\Bigg], \nonumber\\
    & \ \text{where, } \hat{J}_{k} = J+\sum_{r\in \mathbb{Z}^{d}}e^{-ik\cdot r}J_{|r|}
\end{align}

Note that all the $G^{uv}_{k}$ fields can be integrated out, by satisfying the equations of motion, 
\begin{align}
    &G^{uv}_{k}= (-1)^{u+v}\frac{2iF^{uv}_{k}}{\hat{J}_{k}} &&\ \text{for } k\neq 0, \nonumber\\
    &G^{uv}_{0}= \frac{2\pi\gamma}{\hat{J}_{0}}+(-1)^{u+v}\frac{2 iF^{uv}_{0}}{\hat{J}_0} &&\ \text{for } k= 0.
\end{align}
Integrating out the G fields, we get (again dropping constant terms),
\begin{align} \label{eq:action_loc_mom}
    I[F,\mvec{\Omega}]&=  \int dt \int \dbar^{d}k\sum_{\substack{u<v}}\Bigg[-\frac{(-1)^{u+v}}{\hat{J}_k}\left(iF^{uv}_{k} \right)\left(iF^{uv}_{-k} \right)-\frac{2\pi\gamma}{\hat{J}_0}\left(iF^{uv}_{k} \right)\delta(k)+iF_{k}^{uv}\Omega^{uv}_{-k}\Bigg].
\end{align}
Going back to real space, we get,
\begin{align} \label{eq:action_loc_pos}
    I[F,\mvec{\Omega}]&=  \int dt\sum_{\substack{r\\u<v}}\Bigg[-(-1)^{u+v}\sum_{r^{\prime}}\mathcal{J}_{rr^{\prime}}\left(iF^{uv}_{r}\right)\left(iF^{uv}_{r^{\prime}}\right)-\frac{\gamma}{\hat{J}_0}\left(iF^{uv}_{r} \right)+\left(iF_{r}^{uv}\right)\Omega^{uv}_{r}\Bigg],
\end{align}
with an effective real space interaction $\mathcal{J}_{rr^{\prime}}$ given by,
\begin{align}\label{eq:effectiveinteraction}
    \mathcal{J}_{rr^{\prime}}&=\int \dbar^{d} k \frac{e^{ik\cdot (r-r^{\prime})}}{\hat{J}_{k}}.
\end{align}

By identifying the last term in~(\ref{eq:action_loc_pos}) as a partition function for the spins with external fields $iF$, one gets the~Eq.\ref{eq:actionreplicafield} in the main text.

\subsection{Effective real space interaction}\label{appsec:renormalized_locality}

~(\ref{eq:action_loc_pos}) holds for any general number of dimensions. The microscopic interaction between the spins i.e. $J_{rr^{\prime}}$ can be Fourier transformed to give the momentum space interaction $\hat{J}_{k}$ between the $G$ fields. This can then be transformed to an interaction $\mathcal{J}_{rr^{\prime}}$ between the $iF$ fields via ~(\ref{eq:effectiveinteraction}). 

Here we consider two forms of the interaction, nearest neighbor (NN) or power-law interacting (PL). In the main text results are used for the PL case. Including the on-site term, the real space interaction between spins in a d-dimensional lattice is defined as follows, 
\begin{align}
    J_{rr^{\prime}}&=
    \begin{cases}
        J\left(\delta_{r,r^{\prime}}+g\sum_{i}\left(\delta_{r,r^{\prime}+e_{i}}+\delta_{r,r^{\prime}-e_{i}}\right)\right) &\text{ (NN)}\\
        J\left(\delta_{r,r^{\prime}}+g\left(1-\delta_{r,r^{\prime}}\right)\frac{1}{|r-r^{\prime}|^{2\alpha}}\right) &\text{ (PL)}.
    \end{cases}
\end{align}
Here $e_{i}$ is a d-dimensional vector $\{0,0,..,i,..,0\}$ with 1 in the i-th position.

The momentum space interaction is given by its discrete-time Fourier transform,

\begin{align}
    \hat{J}_{k} = \begin{cases}
        J\left(1+2g\sum_{i}\cos{k_{i}}\right) & \text{ (NN)}\\
        J\left(1+g\sum_{s\neq 0}\frac{e^{-i k\cdot s}}{|s|^{2\alpha}}\right) &\text{ (PL)}.
    \end{cases}
\end{align}

To get the effective real space interaction $\mathcal{J}$ between the replica fields $iF$, one has to take the Fourier transform of $\hat{J}_{k}^{-1}$. This can be done exactly for $d=1$.
\subsubsection{d=1}
In $d=1$ we have,

\begin{align}
    \hat{J}_{k} = \begin{cases}
        J\left(1+2g\cos{k}\right) & \text{ (NN)}\\
        J\left(1+g\left( \text{Li}_{2 \alpha }\left(e^{-i k}\right)+\text{Li}_{2 \alpha }\left(e^{i k}\right)\right)\right) &\text{ (PL)},
    \end{cases}
\end{align}
where $\text{Li}_{n}\left(z\right)$  is the Polylogarithm function. For large $\alpha$, $\text{Li}_{2\alpha}(z)\to z$, and $\hat{J}_k$ reduces to the Nearest Neighbor case.

In the NN case, the real space interaction between the replica fields is given by,
\begin{align}
    \mathcal{J}^{\text{NN}}_{rr^{\prime}} = \int_{0}^{2\pi}\frac{dk}{2\pi}\frac{e^{ik(r-r^{\prime})}}{1+2g\cos{k}}.
\end{align}
Changing the variables to $z = e^{ik}$, we get contour integral defined along the unit circle in the complex $z$-plane, with two isolated poles along the negative real axis. Performing the contour integral picks up the pole within the unit circle and one obtains,
\begin{align}
    \mathcal{J}^{\text{NN}}_{rr^{\prime}}=\frac{(-1)^{r-r^{\prime}}}{\sqrt{1-4g^{2}}}e^{-\text{acosh}\frac{1}{2g}|r-r^{\prime}|}.
\end{align}

In the PL case, with the same change of variables, the real space interaction is given by the following contour integral defined along the unit circle,
\begin{align}
     \mathcal{J}^{\text{PL}}_{rr^{\prime}} = \int_{\mathcal{C}}\frac{dz}{2\pi i}\frac{z^{r-r^{\prime}-1}}{1+g\left(\text{Li}_{2\alpha}(z)+\text{Li}_{2\alpha}(1/z)\right)}.
\end{align}
For large enough $g$ (including $g=1$) there is an isolated pole along the negative real axis within the unit circle, and a branch cut due to the $\text{Li}_{2\alpha}(1/z)$ term along the positive real axis $z\in(0,\infty)$. The pole gives an exponential decay like the NN case, while we will show that the branch cut contribution leads to a power law interaction.

Deforming the contour to hug the branch cut, the integral is proportional to the discontinuity along the branch cut ($\int_{0+i\varepsilon}^{1+i\varepsilon}-\int_{0-i\varepsilon}^{1-i\varepsilon}$), which for the function $\text{Li}_{2\alpha}(1/z)$ is proportional to $\log^{2\alpha-1}(1/z)$. Thus we get,
\begin{align}
    \mathcal{J}^{\text{PL}}_{rr^{\prime}} = \text{Pole contribution} -\int_{0}^{1}dz z^{r-r^{\prime}-1} \log^{2\alpha-1}(z) g(z),
\end{align}
where $g(z)$ is a smooth function. For large $|r-r^{\prime}|$ the integrand is heavily suppressed away from $z=1$, so we can change variables $z\sim e^{-w}$, and after dropping the regular terms, we have an integral,
\begin{align*}
    \int_{0}^{\infty}dw e^{-w|r-r^{\prime}|}w^{2\alpha-1} \sim |r-r^{\prime}|^{-2\alpha}, \text{ for }|r-r^{\prime}|\gg 1.
\end{align*}

Along with the pole contribution we thus get the effective real space interaction,

\begin{align}
    \label{eq:jrrplcouplings}
    \mathcal{J}^{\text{PL}}_{rr^{\prime}} \sim (-1)^{r-r^{\prime}}e^{-\mu  |r-r^{\prime}|}-(1-\delta_{rr^{\prime}})\frac{1}{|r-r^{\prime}|^{2\alpha}}.
\end{align}
At large $|r-r^{\prime}|$, $\mathcal{J}_{rr^{\prime}}\sim |r-r^{\prime}|^{-2\alpha}$, as was noted in the main text.

\subsubsection{General d}
In this section we will derive the effective real space interaction in any general dimension, $d$. 

For the NN case, we have, 
\begin{align}
    \mathcal{J}^{\text{NN}}_{rr^{\prime}} = \int\dbar^{d}k\frac{e^{ik\cdot(r-r^{\prime})}}{1+2g\sum_{i}\cos{k_{i}}}, 
\end{align}
which can be expanded around $k_{i} = \pi$. Furthermore, we can consider $r-r^{\prime}$ to be along a particular dimension, say 1. With these manipulations we get, 
\begin{align}
    \mathcal{J}^{\text{NN}}_{rr^{\prime}} &\sim  \int_{-\pi}^{+\pi}\dbar^{d}k\frac{e^{ik_{1}(r-r^{\prime})}}{1+\tilde{g}\sum_{i}k_{i}^{2}} \nonumber \\
    &\sim \int_{0}^{\infty}dk k^{d-2} \int \dbar k_{1}\frac{e^{ik_{1}(r-r^{\prime})}}{1+k_{1}^{2}+k^{2}} \nonumber\\ 
    &\sim \int_{0}^{\infty} \frac{k^{d-2}}{\sqrt{1+k^{2}}}e^{-\sqrt{1+k^{2}}|r-r^{\prime}|} \sim e^{-\mu |r-r^{\prime}|}.
\end{align}

For the PL case, one needs to evaluate,
\begin{align}
    \mathcal{J}^{\text{PL}}_{rr^{\prime}} \sim \int \dbar^{d}k \frac{e^{i k \cdot s}}{1+g\sum_{u\neq 0}\frac{e^{-ik\cdot u}}{|u|^{2\alpha}}}.
\end{align}
Firstly we have,
\begin{align}
    \sum_{u\neq 0}\frac{e^{-ik\cdot u}}{|u|^{2\alpha}} \sim \int^{\prime}\dbar^{d}u \frac{e^{-ik\cdot u}}{|u|^{2\alpha}} \sim k^{2\alpha-d}.
\end{align}
With this, we can expand the function to be Fourier Transformed at small $k$,
\begin{align}
    \mathcal{J}^{\text{PL}}_{rr^{\prime}}\sim \int \dbar^{d}k e^{ik\cdot s}\left(1-g k^{2\alpha-d}\right) \sim \delta(s)-g|s|^{-2\alpha}.
\end{align}

These results generalize~\ref{eq:jrrplcouplings} for any general dimension.

\section{Saddle point analysis of the mean field}
\label{appsec:spinmodelsaddle}

\subsection{Effective Bulk Action and saddle points}

To understand the saddle points of the action $I[\phi,\Theta]$ we first consider the bulk mean-field limit in which the fields are independent of space and time: $\phi_r(t) = \phi, \Theta_r(t) = \Theta$. In this case the propagator $\mathcal{K}_r$ is dominated at long times by the ground state of the r-bit effective Hamiltonian $\phi \sigma_r^x + \Theta \sigma_r^z$, which yields
\begin{equation}
    \sum_r \ln \mathcal{K}_r = \frac{LT}{2} \sqrt{\phi^2 + \Theta^2} + \frac{BLT}{2}.
\end{equation}
By substituting this into Eq. \eqref{eq:2fieldPathIntegralmain} (and again dropping additive constant terms) we obtain the time- and space-independent effective bulk action,

\begin{align}
    \label{eq:rbitbulkaction}
    I_{\text{MF}}[\phi,\Theta] = 
    N T L \left[\mathcal{J}\left(\phi^{2}-3\Theta^{2}\right)-9\left(\Gamma+\frac{1}{9}\right)\Theta-\frac{1}{2}\sqrt{\phi^{2}+\Theta^{2}}\right],
\end{align}
where we have defined $\Gamma \equiv \frac{\gamma}{\hat{J}_0}$ and $\mathcal{J} \equiv 27/16\sum_{s\in \mathbb{Z}}\mathcal{J}_{r(r+s)}$.

\begin{figure}
    \includegraphics[width=0.5\textwidth]{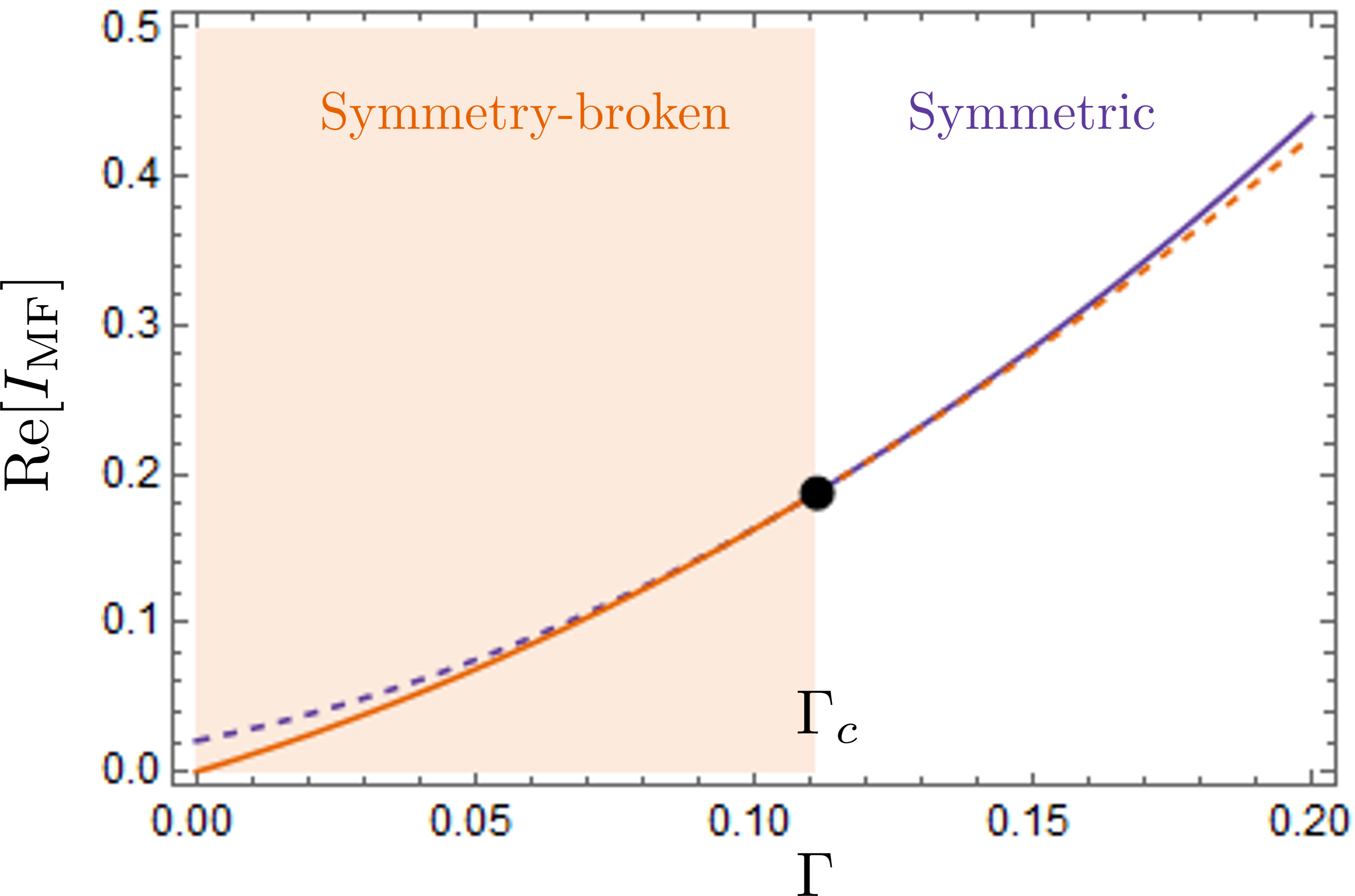}
    \caption{Action cost for symmetric and symmetry-broken phases. Below the critical point $\Gamma < \Gamma_c$, the action cost $\mathrm{Re}[I_{MF}]$ is minimized by the symmetry-broken saddle points (solid orange) relative to the symmetric saddle point (dashed purple). Above the critical point $\Gamma > \Gamma_c$ the path integral is dominated by the symmetric saddle point (solid purple) because the symmetry-broken saddles (dashed orange) are imaginary and therefore do not contribute to the integral (see Fig. \ref{fig:contourplots}).}
    \label{fig:actioncomp}
\end{figure}

Saddle point analysis of this action  \cite{bentsen2021measurementinduced} reveals a phase transition at $\Gamma_{c} = 1/9$ (see Fig.~\ref{fig:actioncomp}). For $\Gamma > \Gamma_c$ the action is dominated by the symmetric saddle point with $\phi^*=0$ and $\Theta^* = -3(\Gamma_c+2 \Gamma) / 4 \mathcal{J}$. By contrast, for $\Gamma < \Gamma_c$, the action is dominated by the symmetry-broken saddle-point where the field $\phi^*$ is non-zero and comes in a pair, $\phi^* \propto \pm \sqrt{\Gamma_{c}-\Gamma}$, while $\Theta^* = -9(\Gamma_c + \Gamma) / 8 \mathcal{J}$. The replica permutation $1\leftrightarrow 3$ is equivalent to $\phi\leftrightarrow-\phi$ symmetry, which is spontaneously broken for $\Gamma < \Gamma_c$.
\begin{figure*}
    \includegraphics[width=0.85\textwidth]{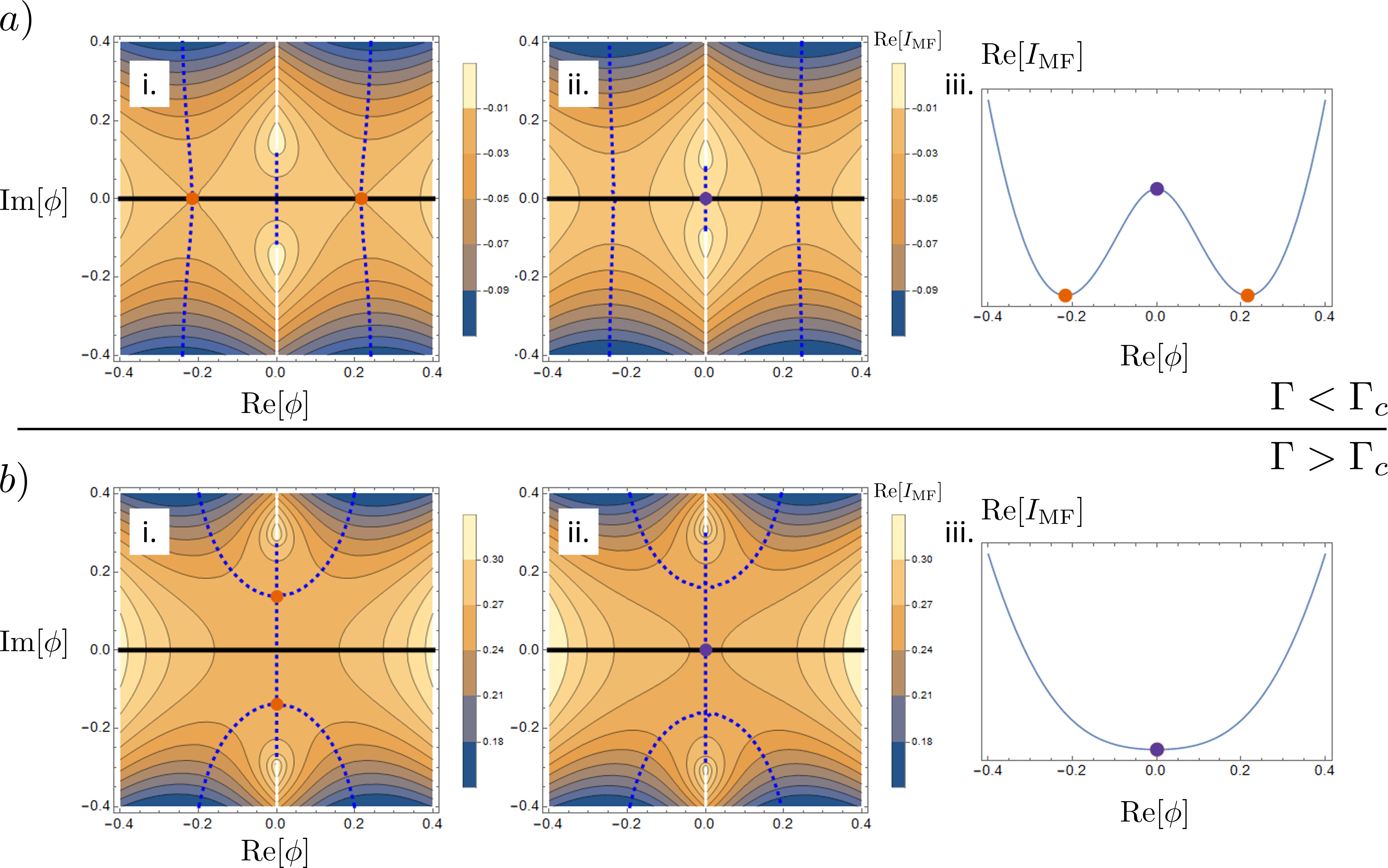}
    \caption{Saddle-point integration contours. (a) Below the critical point, both the symmetry-broken saddle points (i., orange dots) and the symmetric saddle point (ii., purple dot) lie along the real axis of integration (solid black). In this case all three saddle points contribute to the integral, but the symmetry-broken saddle points dominate because they minimize the action $\re{I_{\mathrm{MF}}}$. (b) Above the critical point, the symmetry-broken saddle points (i., orange dots) lie on the imaginary axis and therefore do not contribute to the integral. Therefore the integral is dominated by the symmetric saddle point (ii., purple dot). Dotted blue lines show contours of steepest ascent / descent.}
    \label{fig:contourplots}
\end{figure*}

Note, $\gamma_{c}$ is given by $\Gamma_{c}\hat{J_{0}}\sim \Gamma_{c}\left(1+g\sum_{s\neq 0}|s|^{-2\alpha}\right) \approx \Gamma_{c}\left(1+g \int^{\prime}d^{d}s|s|^{-2\alpha}\right)$. Thus $\gamma_{c}$ diverges when $2\alpha < d$, when the system effectively becomes all-to-all.

\subsection{Effective Field Theory Near Criticality}

In the previous analysis we understood the bulk mean-field physics of the model by ignoring the space- and time- dependence of the fields $\phi_r(t),\Theta_r(t)$. Here we restore the space- and time-dependence and study the model near the critical point. In this limit we can transform the effective r-bit action Eq. \eqref{eq:2fieldPathIntegralmain} into a $\phi^4$ field theory with long-range interactions. In particular, near the critical point the dynamics is governed entirely by the small fluctuations of the symmetry-breaking field $\phi_r(t)$. In the following we expand the action $I[\phi,\Theta]$ in fluctuations of this small parameter to obtain an effective long-range $\phi^4$ field theory for the system.

First we evaluate the leading term of the effective action $I[\phi,\Theta]$ in Eq. \eqref{eq:2fieldPathIntegralmain}, corresponding to the long-range power-law interactions. Substituting $\mathcal{J}_{rr'} = \mathcal{J}_{rr'}^{\mathrm{PL}}$ from Eq. \eqref{eq:jrrplcouplings}, we obtain
\begin{align}
    \sum_{r'}(-1)^{r-r'}e^{-\mu |r-r'|} \phi_r \phi_{r'} &\approx \sum_{q\in \mathbb{Z}}(-1)^{q}e^{-\mu|q|}\phi_r^{2}+\sum_{q\in \mathbb{Z}}(-1)^{q}\frac{q^{2}}{2}e^{-\mu|q|}\phi_r \partial_{r}^{2} \phi_r + O(\phi_r \partial_r^4 \phi_r) \nonumber \\
    & \approx \frac{(1-e^{-\mu})}{(1+e^{-\mu})} \phi_r^{2}- \frac{e^{-2\mu}\left(1-e^{-\mu}\right)}{\left(1+e^{-\mu}\right)^{3}} \phi_r \partial_{r}^{2} \phi_r
\end{align}
and
\begin{align}
    \sum_{r'} (1-\delta_{rr^{\prime}})\frac{1}{|r-r^{\prime}|^{2\alpha}} \phi_r \phi_{r'} \rightarrow \int^{\prime}_{s} \frac{1}{|r-s|^{2\alpha}} \phi_r \phi_{s} 
\end{align}
in the continuum limit, where the delta-function term $1-\delta_{rr'}$ is responsible for the UV cut-off in the continuum integral $\int^{\prime}_{s} = \int_{\mathbb{R}\backslash(r-\varepsilon,r+\varepsilon)}ds$. Here we have assumed that the parameter $\mu$ is large such that the exponential $e^{-\mu \magn{q}}$ decays rapidly and suppresses higher derivative terms e.g. $\phi_r \partial_r^4 \phi_r$. Next, we fix $\Theta_r(t) = \Theta = -3(\Gamma_c+2 \Gamma) / 4 \mathcal{J}$ to its saddle-point value and expand the propagators $\ln \mathcal{K}_r$ in the small parameter $\phi_r(t)$ \cite{bentsen2021measurementinduced}:
\begin{align}
    \ln \mathcal{K}_r &= \ln 2 \cosh{\frac{\Theta T}{2}} +\frac{1}{8}\int dt_{1}dt_{2} \phi_r(t_1) \phi_r(t_2)e^{-\left|\Theta\right|\left|t_1-t_2\right|}-\int dt \frac{\phi_r^{4}}{16\left|\Theta\right|^{3}}+O(\phi_r^{6}) \nonumber \\
    &\approx \int dt \left[\frac{\phi_r \partial_{t}^{2} \phi_r}{4\left|\Theta\right|^{3}}+\frac{\phi_r^{2}}{4\left|\Theta\right|}-\frac{\phi_r^{4}}{16\left|\Theta\right|^{3}}\right] + \mathrm{const}.
\end{align}
Finally, summing the results, taking the continuum limit, and ignoring additive constants, we find the near-critical Landau-Ginzburg effective field theory
\bea 
    \frac{I_{\text{eff}}[\phi]}{N}= \frac{1}{4 \magn{\Theta}^3} \int_{t,r} dt dr \Bigg[ -  \phi_r \left( \partial_{t}^{2} + \beta \partial_r^2 \right) \phi_r -b \int^{\prime}_{s} \frac{\phi_{r}\phi_{s}}{|r-s|^{2\alpha}}-\frac{\delta}{2}\phi_r^{2}+\frac{\  \phi_r^{4}}{4}\Bigg],
\eea
for small fluctuations $\phi_r(t)$ around the critical point. Here we have introduced the numerical coefficients
\begin{align}
    \beta &= 4 \magn{\Theta}^{3} \frac{27 J}{16}\frac{e^{-2\mu}\left(1-e^{-\mu}\right)}{\left(1+e^{-\mu}\right)^{3}} \nonumber \\
    b &= 4 \magn{\Theta}^3 \frac{27 J}{16} \nonumber \\
    \delta &= 8 \magn{\Theta}^{3} \left(\frac{1}{4 \magn{\Theta}} - \frac{27 J}{16} \frac{(1-e^{-\mu})}{(1+e^{-\mu})}\right).
\end{align}
By rescaling space and $b$ we can get rid of the $\beta$ parameter, and end up with the Landau-Ginzburg theory in~Eq. 6 in the main text.

\section{Scaling of entanglement entropy in the symmetry broken  phase of the nearest neighbor model}\label{appsec:symmbroken}
In this section we derive results for $d=1$. For the nearest neighbor case, the power law term in the action is absent and we will be working with the short-range action,
\begin{align}
    I[\phi] &=  \int dt \int dr \left[ -\phi\left(\partial_{t}^{2}+\beta\partial_{r}^{2}\right)\phi-\frac{\delta}{2}\phi^{2}+\frac{\phi^{4}}{4}\right], 
\end{align}
\subsection{Entropy of a maximally mixed initial state}
The SWAP action acts on the whole spatial slice, and we need to only solve for the equations of motion with the time derivative,
\begin{equation}
    \partial_t^{2}\phi = -\delta \phi +\phi^{3}.
\end{equation}
For $\delta<0$, we have an instanton-like solution,
\begin{align}
    \phi^{*}(r,t) = \sqrt{\delta}\tanh\left[\sqrt{\frac{\delta}{2}}\left(t-t_{0}\right)\right].
\end{align}
Hence for $\delta > 0$, we have the following scaling of the quasi-entropy of the full system,
\begin{align}
    \hat{S}^{(2)}_Q\sim N\left(I[\phi^{*}]-I[\sqrt{\delta}]\right) \sim NL\delta^{3/2} + ...
\end{align}

The correction to this term will be given by the fluctuations of the domain wall within periodic boundary condition, which was estimated using Capillary Wave Theory in \cite{li2020statistical}. For $T\gg \sqrt{L}$, this will be given by,
\begin{align}
    -\log \frac{T}{\sqrt{L}} - \log \sqrt{\frac{\beta N \delta^{3/2}}{12\pi}} \sim -\log \frac{T}{\sqrt{L}} + \text{const.}
\end{align}

The expression for the entropy of the entire system $S_{Q}$ is essentially the same for the power-law interacting model, with suitably renormalized $\delta$, 

\begin{equation}
    \delta\to \delta_{\alpha} =\delta +\int^{\prime}ds|r-s|^{-2\alpha}. 
\end{equation}

\subsection{Sub-system entropy}
For a subsystem between the region $r_1$ and $r_2$, the domain wall must be pinned to the future time boundary at $r_1$ and $r_2$. We assume the solution of the fields is of the form, 
\begin{align}
    \phi^{*}(r,t) = \sqrt{\delta}\tanh\left[\sqrt{\frac{\delta}{2}}\left(t-y(r)\right)\right],
\end{align}
where $y(r)$ is the `height' of the domain wall (or equivalently the position of the instanton). Due to the boundary pinning effect, $y(r)$ must be $\epsilon\to0$ at $r=r_1$ and $r=r_2$. The action for the quasi-entropy is given by a functional of $y(r)$, 
\begin{align*}
    I[y(r)] &= I[\phi^{*}]-I[\sqrt{\delta}]\\
    &=\int dt \int dr \delta^{2}\Bigg[\left(1+\beta y'(r)^{2}\right)\text{sech}^{2}\sqrt{\frac{\delta}{2}}(t-y) -\frac{1}{4}\left(3+4\beta y'(r)^{2}\right)\text{sech}^{4}\sqrt{\frac{\delta}{2}}(t-y) \\
    & +\frac{\beta}{\sqrt{2\delta}}y''(r)\text{sech}^{2}\sqrt{\frac{\delta}{2}}(t-y)\text{tanh}\sqrt{\frac{\delta}{2}}(t-y)\Bigg]
\end{align*}
Rescale the time variable,
\begin{align*}
    z = \sqrt{\frac{\delta}{2}}(t-y(r)).
\end{align*}
The limits for z integration are approximately $0\to\infty$.
The action for quasi-entropy is thus given by,
\begin{align}
    I[y(r)] &\sim \delta^{3/2}\int dr \int dz \left[\left(1+\beta y'(r)^{2}\right)\text{sech}^{2}z-\frac{1}{4}\left(3+4\beta y'(r)^{2}\right)\text{sech}^{4}z +\mathcal{O}(y''(r))\right]\nonumber\\
    &\sim \delta^{3/2}\int dr \left(1+\frac{\beta}{6}y'(r)^{2}+\mathcal{O}(y''(r)\right).
\end{align}
We can ignore the $\mathcal{O}(y''(r))$ term as it is irrelevant under RG flow, since $[y] = 1$ and $[y'']=-1$. Thus, the quasi-entropy is thus given by,
\begin{align}
    \hat{S}^{(2)}(A) &= -\ln \int \mathcal{D}(y(r))\exp \left[-N\delta^{3/2}\int_{A}dr\left(1+\frac{\beta}{6}y'(r)^{2}\right)\right]\\
    &= N\delta^{3/2}|A|-\ln \int \mathcal{D}(y(r))\exp \left[-\frac{\beta N\delta^{3/2}}{6}\int_{A}dr\left(y'(r)^{2}\right)\right]
\end{align}
This is exactly the action conjectured by Li and Fisher and solved using Capillary Wave Theory in \cite{li2020statistical}. Using their result, the quasi-entropy is given by,
\begin{align}
    \hat{S}_{A}^{(2)} &\approx N\delta^{3/2}|A|+\frac{3}{2}\ln|A|-\ln\left[\sqrt{\frac{2}{\pi}}\epsilon^{2}\left(\frac{\beta N \delta^{3/2}}{6}\right)^{3/2}\right]
    \\
    &=N\delta^{3/2}|A|+\frac{3}{2}\ln|A| +\text{const...}
\end{align}

The approximations a la Li and Fisher \cite{li2020statistical} behind this are $|A|\ll L$, $\sqrt{A} \ll T$, and for a spatial lattice cut-off, $\epsilon$, such that, $\epsilon \ll \sqrt{|A|/(\beta N \delta^{3/2})}$. Note, we further have our large N approximation, $T\lesssim \text{poly}( N)$. These can all be satisfied in the regime, $\sqrt{A}\ll T \lesssim \text{poly}(N)$. The logarithmic correction term is $1/N$ suppressed in our model.

For general long-range interactions, the simple capillary wave picture fails, as the gradient expansion doesn't converge for $2\alpha < 3$ (for the effectively short-ranged case, the story is the same for the nearest neighbor model). In this case, we can't compute the $\mathcal{O}(N^{0})$ correction to $\hat{S}_{A}^{(2)}$. However, compared to the nearest neighbor case, long-range interaction adds an energy cost to the domain wall, and the entropy, as captured in Eq.~\ref{eq:entropyInteracting} in the main text.

\section{Monitored Brownian SYK: the interacting case}
\label{appsec:syk}

In this section we discuss the Landau Ginzburg theory of the quasi-entropy for the interacting case $ \tilde U > 0$. 

The parameter $\lambda$ is determined by~(\ref{eq:lambda}).
For small $\tilde U$, $\lambda = \sqrt{1 - \tilde \gamma^2} [ 1 + \tilde \gamma^2 ( 1 - \tilde \gamma^2)^{q/2-2} \tilde U + O(\tilde U^2) ]$
is well defined when $\tilde \gamma < 1$, and vanishes continuous as $\tilde\gamma \rightarrow 1$.
In the following we will focus on $q=4$, while our results are true for generic $q$.
At the critical point, 
\bea \label{eq:x_condition}
    \lambda^2 \big((2\tilde U-1)+(\tilde U^2-2 \tilde U) \lambda^2 - \tilde U^2 \lambda^4 \big) =0,
\eea
which shows that for $\tilde U > 1/2$ there are two degenerate distinct physical solutions indicating a discontinuous jump.
Thus, the condition for a continuous transition is $ 2 U < \hat J$.
On the other hand, when $\gamma \ge \hat J$, the solution is same as the noninteracting case~(\ref{eq:SYKq_saddle}) at $\tilde\gamma \ge 1$.

What symmetry out of $O(2) \times O(2)$ is preserved for $\tilde U > 0$? It is easy to show that the symmetry reduces to $C_4 \times C_4$, satisfying the condition $((O^{-1})^{uu'})^{q/2} S^{u'v'} (O^{v'v})^{q/2} = S^{uv} $.
The generator is still given by $\gamma_{(13)}$ and $\gamma_{(24)}$ but the rotation angle is restricted to multiples of $\pi/2$.
The relative rotation symmetry is spontaneously broken by nonzero $\lambda$ in~(\ref{eq:SYKq_saddle}) when $\gamma < J$. Namely, $\lambda$ serves as an order parameter of the $C_4$ symmetry breaking transition.
With a slight modification that replaces $\cos k$ to $\epsilon_k$ and $J$ to $\hat J$, the Landau-Ginzburg effective theory reads
\bea 
\frac{I_{\text{eff}}}{N} &=& \frac12 \sum_{i=1,2; k} \int_{\Omega}  \left( \frac{\Omega^2}{\gamma} + \hat J(1- \epsilon_k) \right) |\phi_{i,k}(\Omega)|^2 +  \sum_r \int_{t} \left( \frac{\gamma- \hat J}2 \vec \phi_{r}^2 + \frac{\gamma}8 \vec \phi_{r}^4  - \frac{U}4 ( \phi_{1,r}^4 + \phi_{2,r}^4 ) \right), 
\eea
where $\phi_1 = \delta G^{12} + \delta G^{34}$ and $\phi_2 = \delta G^{14} + \delta G^{23}$ transform like a vector under the relative $C_4$ rotation. (In this case we have $\delta G^{12}_{RR} = \delta G^{12}_{LL}$, $\delta G^{34}_{RR} = \delta G^{34}_{LL}$, $\delta G^{14}_{RR} = \delta G^{14}_{LL}$, $\delta G^{23}_{RR} = \delta G^{23}_{LL}$. So we omit the subscript of the left and right chains.)
This theory features a second order transition if $2U < \gamma $, and a first order one if $2U>\gamma$, consistent with the analysis~(\ref{eq:x_condition}) of the saddle-point solution at the transition $\gamma= \hat J$. 

\end{appendix}

\end{document}